\documentclass[12pt]{iopart}

\usepackage{graphicx}
\begin{document}

\title{Nonmagnetic impurity perturbation to the quasi-two-dimensional quantum helimagnet LiCu$_2$O$_2$}

\author{H C Hsu$^{1,2}$, J -Y Lin$^3$, W L Lee$^4$, M -W Chu$^1$, T Imai$^5,^6$, Y J Kao$^7$, C D Hu$^7$, H L Liu$^2$, and F C Chou$^{1,8}$}

\address{
$^1$Center for Condensed Matter Sciences, National Taiwan University, Taipei 10617, Taiwan}
\address{
$^2$Department of Physics, National Taiwan Normal University, Taipei 116, Taiwan}
\address{
$^3$Department of Physics, National Chiao Tung University, Hsinchu 30076, Taiwan}
\address{
$^4$Institute of Physics, Academia Sinica,Taipei 11529, Taiwan}
\address{
$^5$Department of Physics, McMaster University, Ontario L8S 4M1, Canada}
\address{
$^{6}$Canadian Institute for Advanced Research, Toronto, Ontario M5G1Z8, Canada}
\address{
$^7$Department of Physics, National Taiwan University, Taipei 10617, Taiwan}
\address{
$^8$National Synchrotron Radiation Research Center, Hsinchu 30076, Taiwan}

\ead{fcchou@ntu.edu.tw}

\begin{abstract}
A complete phase diagram of Zn substituted quantum quasi-two-dimensional helimagnet LiCu$_2$O$_2$ has been presented.  Helical ordering transition temperature (\emph{T$_h$}) of the original LiCu$_2$O$_2$ follows finite size scaling for less than $\sim$ 5.5$\%$ Zn substitution, which implies the existence of finite helimagnetic domains with domain boundaries formed with nearly isolated spins.  Higher Zn substitution $\geq$ 5.5$\%$ quenches the long-range helical ordering and introduces an intriguing Zn level dependent magnetic phase transition with slight thermal hysteresis and a universal quadratic field dependence for \emph{T$_c$}(\emph{z} $<$ 0.055,H).  The magnetic coupling constants of nearest-neighbor (nn) \emph{J$_1$} and  next-nearest-neighbor (nnn) \emph{J$_2$} ($\alpha$=\emph{J$_2$/J$_1$}) are extracted from high temperature series expansion (HTSE) fitting and N=16 finite chain exact diagonalization simulation.  We have also provided evidence of direct correlation between long-range helical spin ordering and the magnitude of electric polarization in this spin driven multiferroic material.

\end{abstract}

%\noindent{\it Keywords}: LiCu$_2$O$_2$; multiferroics; dimerization
\maketitle

\section{Introduction}

%most general introduction to LiCu2O2
LiCu$_2$O$_2$ is a complex spin-driven multiferroics which shows spontaneous electric polarization below the spin spiral ordering temperature near 22 K.\cite{Park2007}  The noncollinear spiral spin structure allows inversion symmetry breaking and manifested a finite electric polarization. The nature of helimagnetic ordering and its correlation to the existence of electric polarization has been the central topic in the research of multiferroics.  In particular, microscopic model based on spin supercurrent or inverse Dzyaloshinskii-Moriya interaction provided the theoretical ground for the generation of electric polarization \textbf{P} within  noncollinear magnets, i.e., \textbf{P} $\propto$ \textbf{Q} $\times$ (\textbf{S$_i$} $\times$ \textbf{S$_j$}), where \textbf{Q} is the cycloidal vector pointed from neighboring spin sites $i$ to $j$.\cite{Katsura2005, Seki2008}  Incommensurate helimagnetic ordering has been found in LiCu$_2$O$_2$ and accompanies weak electric polarization of multiferroic nature, although conflicting results on the cycloidal spin plane assignment and disagreement between experimental ferroelectric properties and spin-current model predictions remain.\cite{Park2007, Seki2008, Masuda2004, Fang2009, Mihaly2006, Moskvin2009}

\begin{figure}
\begin{center}
\includegraphics[width=5in]{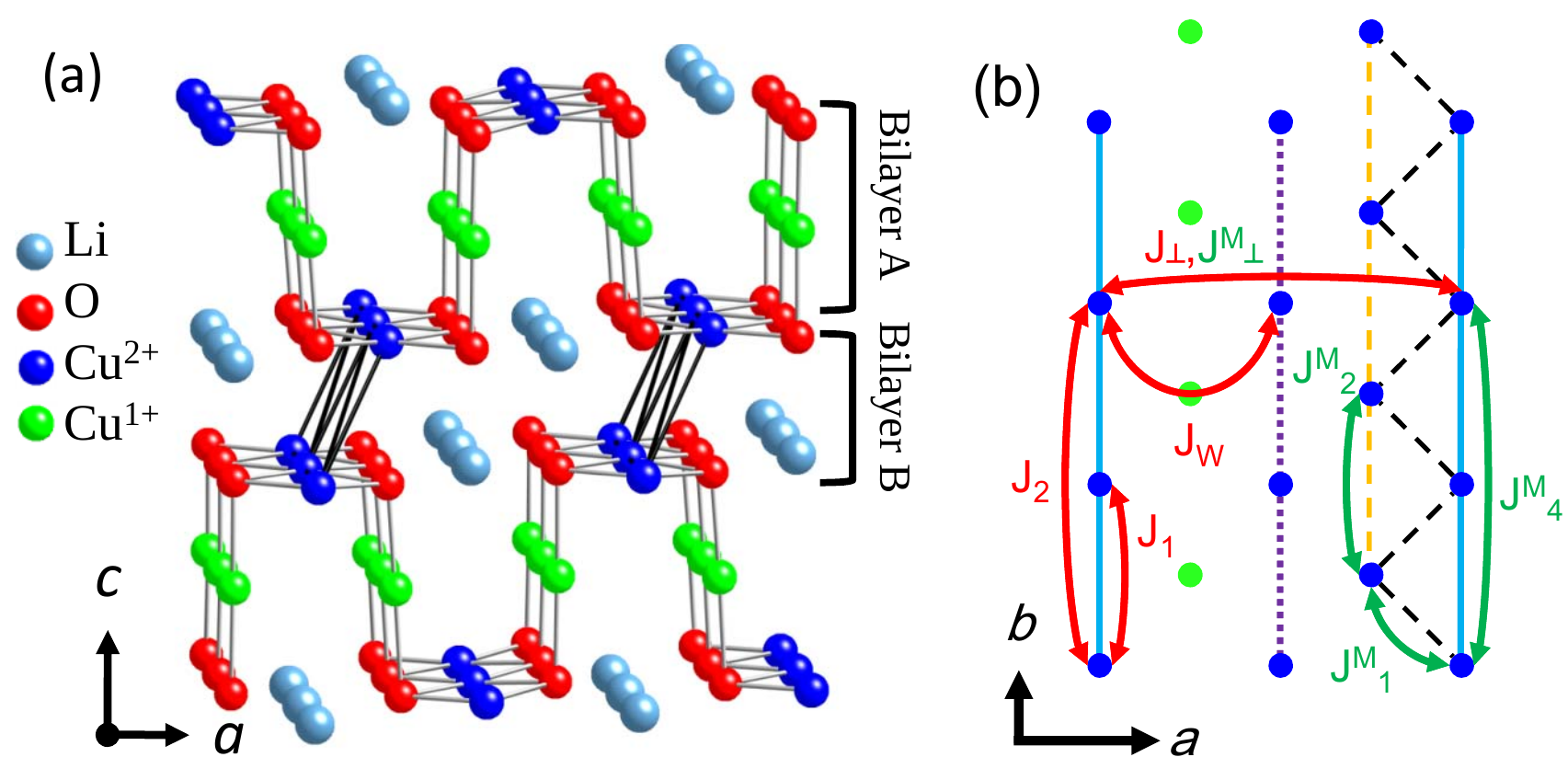}
\end{center}
\caption{\label{fig:fig01}(Color online) (a) Crystal structure of LiCu$_2$O$_2$.  (b) Magnetic couplings described by linear chain model and zigzag chain model.  The purple dot line represents the upper chain within layer A. The orange dashed line represents the lower chain within bilayer B.  Linear chain model is adapted from Gippius \emph{et al.} \cite{Gippius2004} and zigzag chain model is adapted from Masuda \emph{et al.} \cite{Masuda2005} distinguished with superscript M.  Note the \emph{J$_W$} is defined as superexchange coupling between upper and lower Cu$^{2+}$ in bilayer A bridged by the zigzag shape channel through Cu$^+$.}
\end{figure}

%added introduction to the LiCu2O2 study history
LiCu$_2$O$_2$ has a Cu-O chain structure formed with edge-sharing CuO$_4$ plaquettes in the $\emph{ab}$ plane while these chains are connected through CuO$_2$ dumbbells along the \emph{c} direction as shown in figure \ref{fig:fig01}.\cite{Berger1991}  This compound is uniquely composed of nearly equal amount of Cu$^{+}$ and Cu$^{2+}$, with Cu$^{+}$ sitting in the O-Cu-O dumbbell (along the \emph{c} direction) to connect the nearest-neighbor edge-sharing spin chains.  Undoped LiCu$_2$O$_2$ shows a helical spin ordering below $\sim$ 22 K, as indicated by the \emph{d$\chi$/dT} peaks, which has been verified by magnetic neutron scattering studies before.\cite{Seki2008, Masuda2004}  Two \emph{d$\chi$/dT} peaks occur near $\sim$ 22 and 24 K, which are anisotropic and more pronounced for magnetic field applied $\|$ \emph{ab} and $\|$ \emph{c}, respectively.  The anisotropic magnetic correlation lengths $\xi_{ab}$ and $\xi_c$ have been extracted using synchrotron resonant soft x-ray magnetic scattering.  It  is suggested that the transitions near $\sim$ 24 and 22 K correspond to the quasi-two-dimensional short-range and the following three-dimensional long-range orderings respectively.\cite{Seki2008, Huang2008}  Li deficiency is proposed to introduce spin-1/2 Cu$^{2+}$ to the bridging spinless Cu$^+$ site and the helimagnetic ordering temperature is reduced only slightly from $\sim$ 22 K to 20 K for samples nearly free of Li deficiency.\cite{Hsu2008}

%Figure 1

%introduction to the magnetic correlations
Many early works attempted to describe LiCu$_2$O$_2$ as an alternating Heisenberg antiferromagnetic chain and applied Bonner-Fisher fitting to the magnetic susceptibility for \emph{T} $>$ \emph{T$_h$}; however, the fitting is generally unsatisfactory.\cite{Vorotynov1998, Fritschij1998}   Later nearest-neighbor exchange coupling \emph{J$_1$} is found to be ferromagnetic and \emph{J$_2$} to be antiferromagnetic, based both on LDA calculation and neutron scattering spin wave analysis, which satisfies the classical spin spiral ordering requirement of $|$J$_2$/J$_1$$|$ $>$ $\frac{1}{4}$ as verified experimentally.\cite{Gippius2004, Masuda2005}  Quantum fluctuation must sustain in the spin spiral ordered state to account for the discrepancy found in polarized neutron scattering intensity for this frustrated low dimensional spin 1/2 system.\cite{Seki2008}

\begin{figure}
\begin{center}
\includegraphics[width=3.5in]{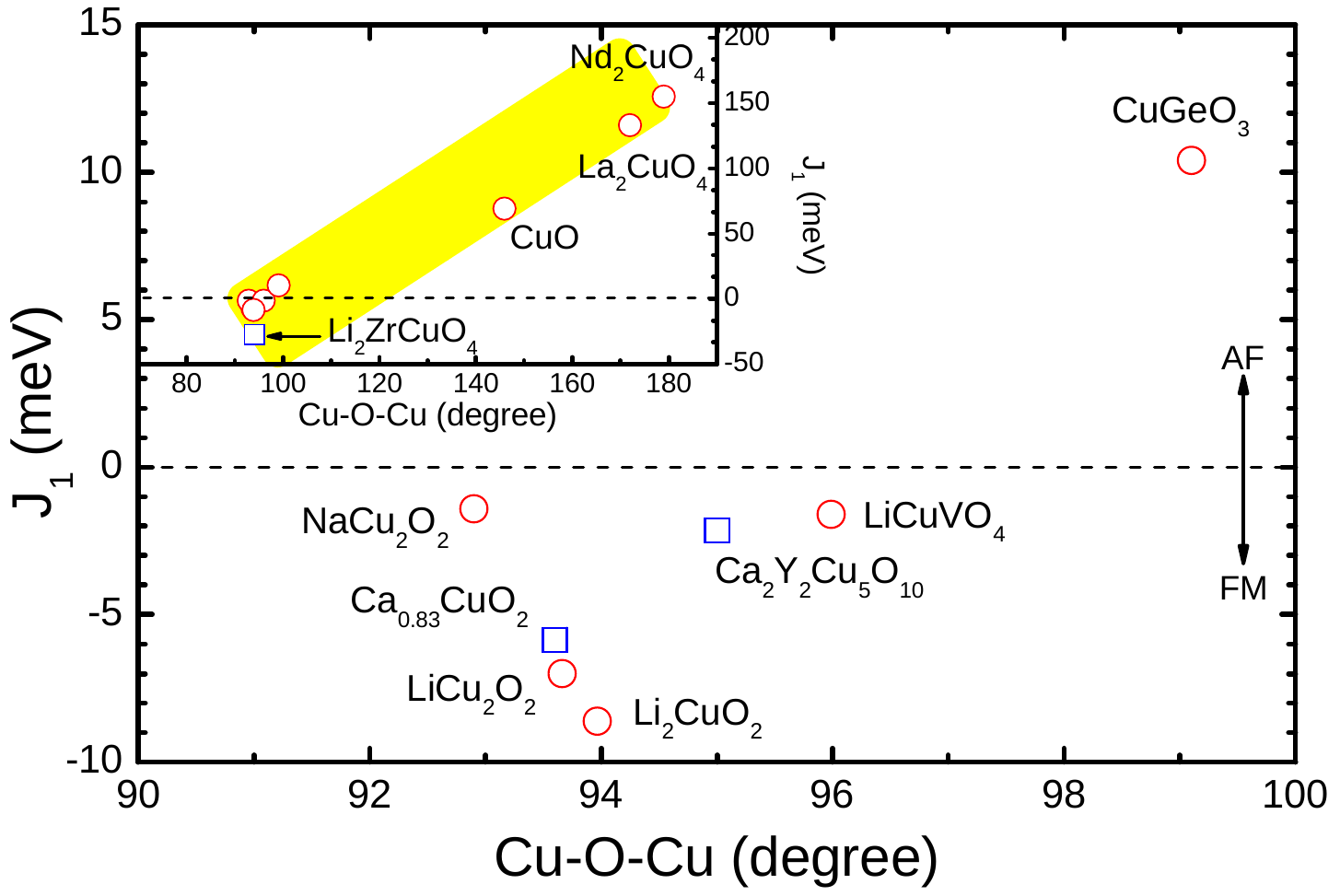}
\end{center}
\caption{\label{fig:fig02}(Color online) Nearest neighbor coupling \emph{J$_1$} versus Cu-O-Cu bond angles summarized for some samples with edge-sharing CuO$_2$ chains, adapted from references \cite{Mizuno1998, Kimura2008}.  The inset shows values up to 180$^\circ$ for mostly two-dimensional layered structure and the main plot amplifies data near 90$^\circ$ region for the chain compounds.  All \emph{J$_1$}'s are obtained from neutron scattering experimental studies, except those from susceptibility fitting and labeled as empty square in blue.}
\end{figure}

If we assign the shortest Cu-Cu distance within the edge-sharing chain to be the \emph{J$_1$}, its Cu-O-Cu bond angle of $\sim$ 94$^\circ$ is expected to be ferromagnetic interaction, following the Goodenough-Kanamori-Anderson rule, as summarized in figure \ref{fig:fig02}.\cite{Mizuno1998, Berger1992, Kimura2008}  There have been two major views on assigning spin chains and its interchain coupling before, one simply viewing the edge-shared CuO$_4$ network as linear spin chains shown in figure \ref{fig:fig01}(b),\cite{Park2007, Drechsler2005} and the other connecting the Cu between linear CuO$_2$ chains to form zigzag chain described by the double layer-B as shown in figure \ref{fig:fig01}(b).\cite{Masuda2004}   Zigzag chain description is only possible when non-negligible coupling \emph{J$^{M}$$_1$} defined in figure \ref{fig:fig01}(b) exists;\cite{Masuda2004} however, LDA calculations and alternative neutron scattering spin wave analysis suggest it to be negligibly small.\cite{Gippius2004, Masuda2005}  On the other hand, based on the model in figure \ref{fig:fig01}(b), the interchain coupling between linear chains through the  Cu-O-Cu superexchange J$_{\bot}$ has been estimated to be comparable to \emph{J$_2$} based on LDA calculations.  In particular, Wannier function modified Hubbard model suggests strong orbital overlap between oxygen and the bridged Cu$^+$ ions, as the \emph{J$_W$} defined in figure \ref{fig:fig01}(b).\cite{Mazurenko2007}  These two kinds of structural views led to different assignment of \emph{J$_1$}, \emph{J$_2$} and interchain coupling \emph{J$_\bot$}, but the discrepancy has been resolved based on neutron scattering spin wave analysis of alternative fittings.\cite{Masuda2005}  Herein we use the linear chain model with non-negligible interchain coupling in the following description and define \emph{J$_1$}, \emph{J$_2$}, $\alpha$=\emph{J$_2$/J$_1$} and interchain coupling \emph{J$_\bot$} based on the model shown in figure \ref{fig:fig01}(b).  When \emph{J$_\bot$} is non-negligible as predicted from both calculation and experiment, \emph{J$_\bot$}  (or \emph{J$_W$}) would connect linear chains equally from both sides along the \emph{a} direction and convert the system from one-dimensional to quasi-two-dimensional.  In fact, such two-dimensional view has been confirmed from its renormalized classical behavior of short-range in-plane magnetic correlation length based on resonant soft x-ray scattering measurement results.\cite{Huang2008}

\begin{figure}
\begin{center}
\includegraphics[width=3.5in]{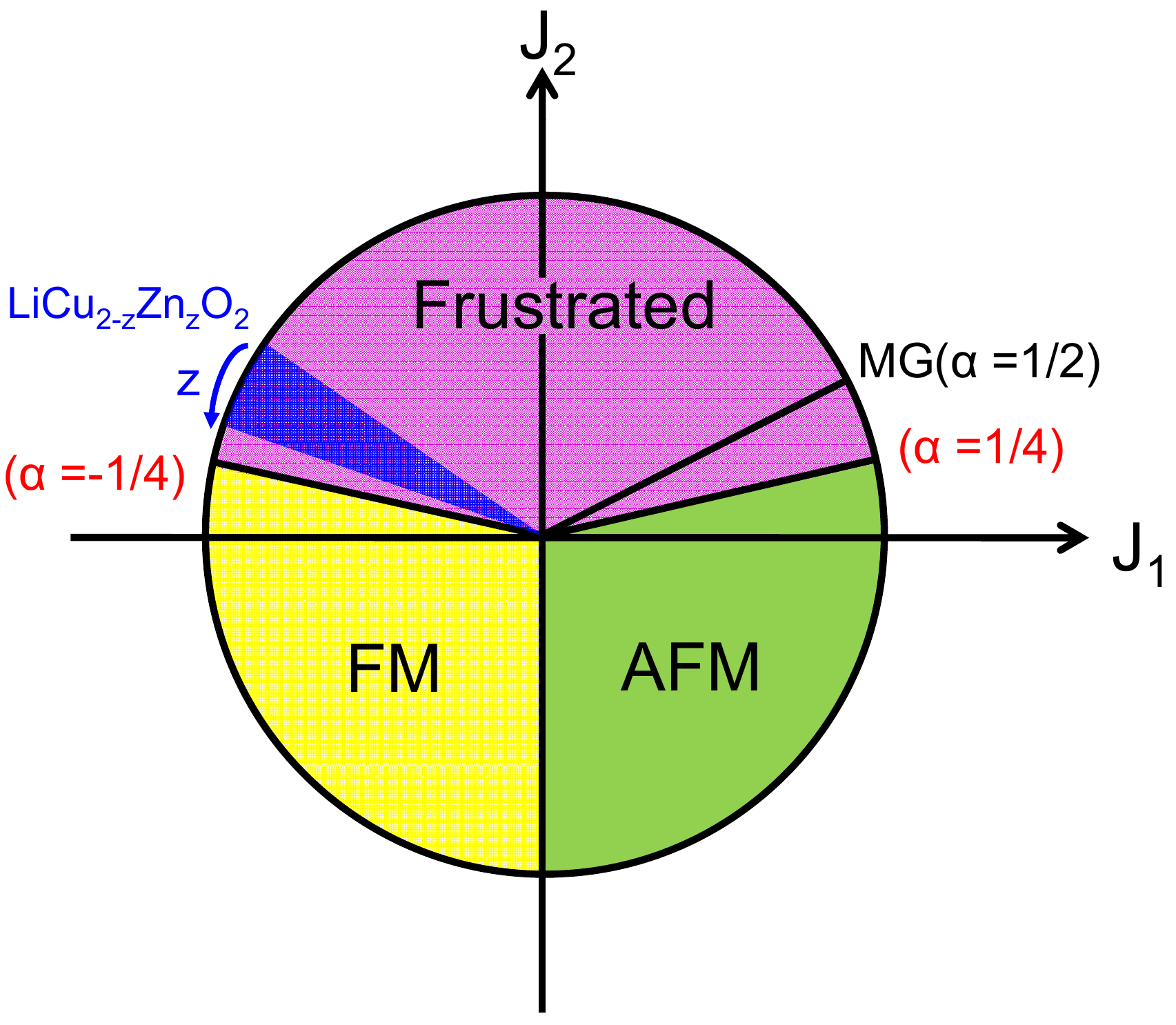}
\end{center}
\caption{\label{fig:fig03}(Color online) \emph{T} = 0 phase diagram of one-dimensional quantum spin 1/2 Heisenberg system with nearest-neighbor \emph{J$_1$} and next-nearest-neighbor \emph{J$_2$} isotropic exchange constants. The Hamiltonian is defined in equation \ref{eq:one}.  Figure is reproduced following Bursill $\textit{et al.}$\cite{Bursill1995}}
\end{figure}

%introduction to the J1/J2 phase diagram
Isotropic spin-1/2 Heisenberg antiferromagnetic chain has been the most widely studied model since the introduction of the Bethe ansatz in 1931.\cite{Bethe1931}  There have been considerable theoretical works on the spin-1/2 chain system with both nearest-neighbor (nn) and next-nearest-neighbor (nnn) of antiferromagnetic interactions as described by the Hamiltonian of
\begin{equation}
H = J_{1} \sum (\textbf{s}_i \cdot \textbf{s}_{i+1} ) + J_{2} \sum (\textbf{s}_i \cdot \textbf{s}_{i+2} ).
\label{eq:one}
\end{equation}
\noindent However, studies on chains with ferromagnetic \emph{J$_1$} and antiferromagnetic \emph{J$_2$} interactions have been rare, and it was only until recently that great interest was re-focused on the quantum spin system with emerging electric polarization concomitant of helical spin ordering, such as LiCu$_2$O$_2$, LiCuVO$_4$ and Li$_2$ZrCuO$_4$ as summarized in figure \ref{fig:fig02}.\cite{Park2007, Masuda2004, Enderle2005, Drechsler2007}  Helimagnetic spin ordering for a spin chain occurs when frustrating \emph{J$_1$} and \emph{J$_2$} satisfy a unique condition of $|$J$_2$/J$_1$$|$ $>$ $\frac{1}{4}$ classically.   We may also view spin periodicity increases from two (for antiferromagnetic ordering) to infinity (for ferromagnetic ordering) with helical ordering of intermediate commensurate or incommensurate periodicity in between in a classical picture, as described in figure \ref{fig:fig03} by its pitch angle $\theta$ = $\arccos$(-\emph{J$_1$/4J$_2$}) growing from zero (\emph{J$_2$/J$_1$}=-1/4) to $\pi$  (\emph{J$_2$/J$_1$}= 1/4).  On the other hand, finite chain calculation on one-dimensional quantum spin-1/2 Heisenberg system suggests the occurrence of a gapped two fold spin singlet (dimer) ground state at \emph{J$^{AF}_{nnn}$/J$^{AF}_{nn}$}=1/2 (called Majumdar-Ghosh, MG, point) and a first order phase transition at \emph{J$_2$/J$_1$} = -1/4 (called FF point) before entering the ferromagnetic regime.\cite{Bursill1995}  The critical ratio of \emph{J$^{AF}_{2}$/J$^{AF}_{1}$} to have a gapped phase from the gapless "spin liquid" (\emph{J$_2$}=0) state for the spin 1/2 antiferromagnetic Heisenberg chain system has been estimated numerically to be $\sim$0.2411.\cite{Okamoto1992}  Finite chain numerical calculations also suggest the occurrence of complex periodicity when approaching the FF point.\cite{Bursill1995} There are very few quantum spin chain systems that sit within the second quadrant of the phase diagram shown in figure \ref{fig:fig03}, except that the Li$_2$ZrCuO$_4$ is claimed to be the one closest to the ferromagnetic critical point,\cite{Drechsler2007, Schmitt2009} LiCu$_2$O$_2$ of incommensurate helical spin ordering with confirmed ferromagnetic \emph{J$_1$} could be the best candidate so far to explore this region.\cite{Enderle2005}

%introduction to the Zn substitution in QM 1d and 2d
Nonmagnetic ion substitution into the low dimensional quantum spin system has been proved to be not just a simple dilution effect.  For example, analytic and quantum Monte Carlo studies on nonmagnetic ion substitution to the Cu site in two-dimensional antiferromagnet La$_2$CuO$_4$ is shown to induce substantial frustrating interactions, approximately $\geq$ 0.4 \emph{J} per impurity, and can resolve the discrepancy between the experimental and fittings by previous theoretical models.\cite{Liu2009}  More interestingly, Zn substitution of Cu in the quantum spin chain CuGeO$_3$ induces an unusual antiferromagnetic ordering coexisting with the spin-Peierls (SP) state of spin-lattice dimerization.\cite{Oseroff1995, Hase1996}  The surprising finding of the coexisting antiferromagnetic and SP states (dimerized antiferromagnetic ordering) is different from the conventional uniform antiferromagnetic state.  It is shown that staggered moments can be introduced and with large spatial inhomogeneity in the ordered moment size (correlation length $\xi \sim$ 10\textbf{a}) through spin chain perturbation.\cite{Fukuyama1996, Kojima1997}  Clearly nonmagnetic spin zero perturbation is an effective way to probe the mysterious  low dimensional quantum spin system.

%early claim in this paper
Herein, we report another surprising finding on Zn substituted quasi-two-dimensional quantum spin-1/2 system LiCu$_2$O$_2$.  Below $\sim$ 5.5 $\%$ Zn substitution, helical ordering temperature \emph{T$_h$(x)} reduces as a function of Zn level and follows an interesting finite size power law, which implies an intriguing domain formation.  Novel phase transition below $\sim$ 20 K is found by $\geq$ 5.5 $\%$ Zn per CuO$_2$ chain substitution.  Magnetic susceptibility measurement results suggest that a new phase transition emerges with a character of spin gap opening with a magnetic field dependence which is similar to that found in a spin-Peierls phase transition.  We tentatively propose one spin dimer model which is consistent to the available magnetic susceptibility, specific heat and NMR measurement results, a satisfactory model fitting and simulation is given as well.   The spin gap size is found to be roughly proportional to the Zn substitution level, and these isolated dimers are suggested to sit on a background of frustrated spins in the finite spin chains cut short by the Zn defects.

\section{Experimental details}

%Experimental method description
A complete series of single crystal LiCu$_{2-z}$Zn$_z$O$_2$ with \emph{z} $\sim$ 0-0.10 have been grown using traveling solvent floating zone (FZ) method.  The grown crystals with chemical analysis are summarized in table \ref{tab:tableI}.   Feed rod of nominal Zn levels of 0, 1, 3, 4, 4.5, 5, 10 and 15$\%$ are prepared through solid state reaction route starting from Li$_{2}$CO$_{3}$ : CuO : ZnO mixture of molar ratio 1.2 : 4-\emph{z} : \emph{z}, where 20 $\%$ excess Li is added to the initial to compensate for the Li vapor loss.  The feed rod is annealed at 750 $^\circ$C after the thoroughly ground powder mixture is treated at 850 $^\circ$C under O$_{2}$ flow for 12 hours each for one time with intermediate grinding.  To reduce further Li loss during the molten stage and to prevent Cu$^{2+}$-rich Li$_2$CuO$_2$ impurity phase formation, high pressure ($\sim$ 0.64 MPa) argon atmosphere is used  during crystal pulling and the pulling rate is maintained at 3 mm/hr with 20 rpm feed/seed rods in counter rotation.  Li content of the Zn-free crystal has been determined using combined thermal gravimetric analysis (TGA) and iodometric titration methods as reported earlier,\cite{Hsu2008} and the Cu and Zn contents have been determined using combined Inductive Coupled Plasma (ICP) and Electron Probe Microanalysis (EPMA).  While it is impossible to determine the Li content accurately with combined TGA and titration methods for the Zn substituted samples, and ICP technique alone cannot provide accurate Li content within 10 $\%$ error, we assume that Li content does not change under the same high pressure Argon atmosphere growth condition and should be maintained near 0.87$\pm$0.03 according to our previous conclusions on the pure Li$_x$Cu$_2$O$_2$ crystal.\cite{Hsu2008}  Details of FZ growth and chemical analysis of Li$_x$Cu$_2$O$_2$ single crystal have been reported previously; however, we describe Li content to be 1 per formula for simplicity in this paper.  The homogeneity issue of Zn substitution is examined with EPMA mapping along the growth direction within 2 $\%$ error within 8 mm and the magnetic phase transition temperature is single and with sharp transition width represented by \emph{d$\chi$(T)/dT} peak near phase transition.

\begin{table}
\begin{center}
\caption{\label{tab:tableI} Zn concentrations determined by EPMA and ICP for LiCu$_{2-z}$Zn$_z$O$_{2}$ crystals studied.}
\begin{tabular}{ccccc}
 \hline\hline
sample ID & \emph{z} (nominal)& \emph{z} (EPMA) & \emph{z} (ICP) \\
 \hline
Zn0 & 0      & NA        &  NA \\
Zn1 & 0.01   & 0.016(3)  & 0.008(4) \\
Zn2 & 0.03   & 0.034(2)  & 0.023(7) \\
Zn3 & 0.04   & 0.038(2)  &          \\
Zn4 & 0.045  & 0.041(1)  &          \\
Zn5 & 0.045  & 0.052(1)  &          \\
Zn6 & 0.05   & 0.058(2)  & 0.050(6) \\
Zn7 & 0.10   & 0.081(3)  & 0.080(5) \\
Zn8 & 0.15  & 0.109(3) & 0.108(4) \\
 \hline\hline
\end{tabular}
\end{center}
\end{table}

Magnetic susceptibilities are measured with SQUID magnetometer (Quantum Design MPMS-XL) with a magnetic field of 1 kOe applied along and perpendicular to the $ab$ plane.  The helimagnetic spin ordering cannot be identified clearly from magnetic susceptibilities, where \emph{$\chi$(T)} drops smoothly at the transition temperature but can only be identified clearly by its derivative \emph{d$\chi$/dT} peaks.  Part of the x-ray structure data were taken using synchrotron x-ray facility NSRRC in Taiwan.   For dielectric measurement, the crystals were shaped into a thin disk with thickness approximately 50-100 $\mu$m. The top and bottom surfaces were then covered by silver epoxy as electrodes. The dielectric constant was measured by a high-precision capacitance bridge at 1 kHz. At room temperature, the dielectric loss is less than 1,000 nano-Siemens (nS) and decreases at lower temperature reaching 0.02 nS below 30 K. The electrical polarization is determined by integration of the pyroelectric current which is detected at a rate of 0.167 K/s after cooling the samples in a poling field of 1200 kV/m from 50 K.  $^{7}$Li NMR measurements were carried out based on standard pulsed NMR techniques.  Field-swept NMR lineshapes were measured by integrating the spin echo signal at a fixed frequency of $f = 33.095$~MHz while sweeping the magnetic field.  Spin-lattice relaxation rate $1/T_{1}$ was measured by applying an inversion pulse prior to the spin echo sequence with varied delay times.

\section{Results and Discussions}

\begin{figure}
\begin{center}
\includegraphics[width=3.5in]{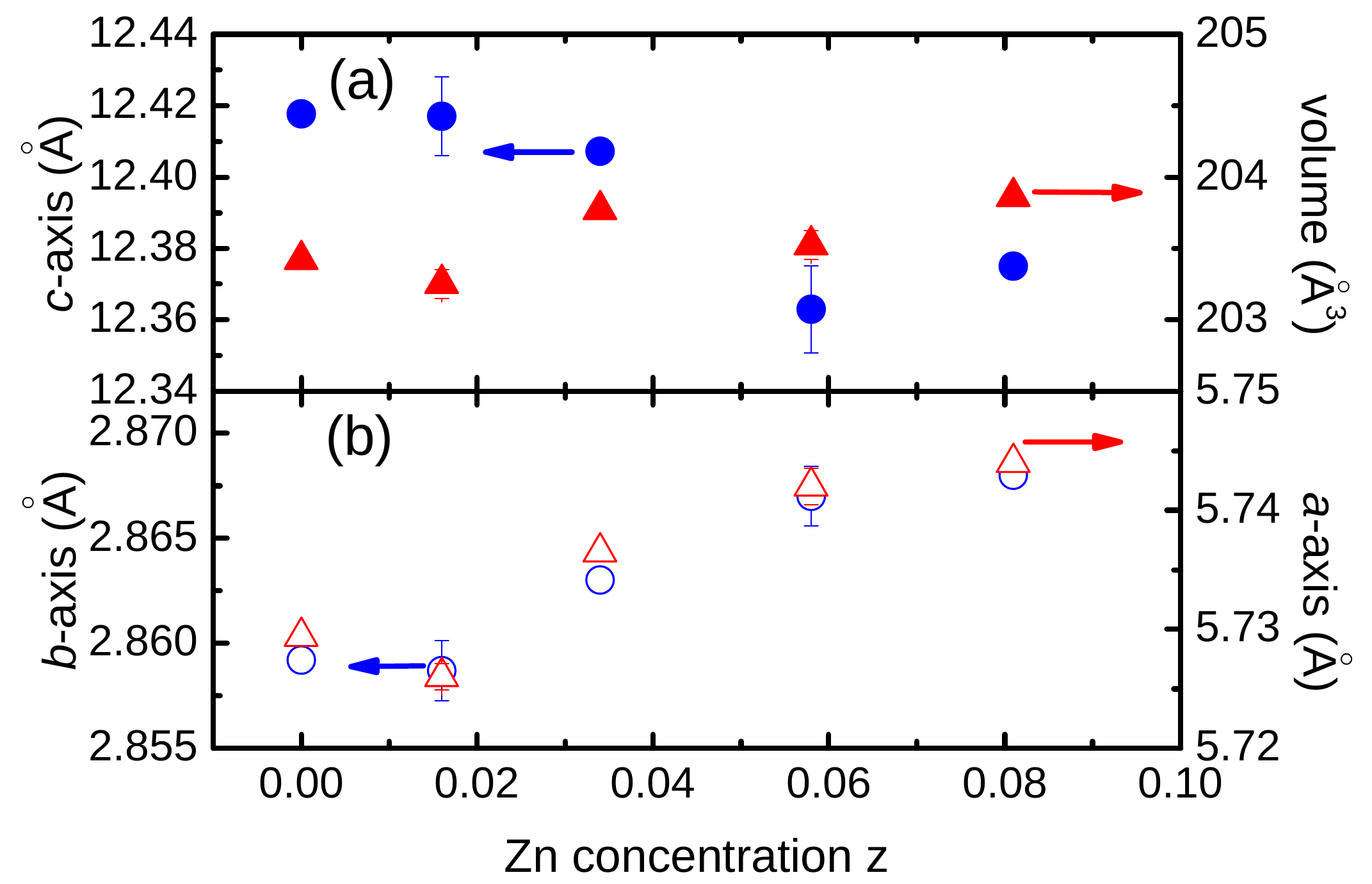}
\end{center}
\caption{\label{fig:fig04}(Color online)  Lattice parameters of LiCu$_{2-z}$Zn$_{z}$O$_{2}$ with various Zn substitution levels.   (a) The c axis is shown in filled circle and volume is shown in filled triangle, and the circle and triangle in (b) represent b- and a-axes, respectively.}
\end{figure}

%Zn substitution consideration
Nonmagnetic Zn ion substitution to the spin 1/2 Cu$^{2+}$ site within edge-sharing CuO$_2$ linear chain is expected to disrupt \emph{J$_1$} effectively and to cut down the infinite chain into even and odd finite chains, although quantum fluctuation, next nearest neighbor \emph{J$_2$}, and interchain coupling \emph{J$_\bot$} could still bring up unexpected results.  Although there are two different structural sites of Cu in LiCu$_2$O$_2$ as shown in figure \ref{fig:fig01}, nonmagnetic Zn$^{2+}$ should substitute the Cu$^{2+}$ site only from  considerations of both valence and ionic radius.   The ionic radius of Zn$^{2+}$ ion (0.74 {\AA} \emph{CN}=4) is similar to that of Cu$^{2+}$ ion (0.71 {\AA} \emph{CN}=4) for square planar coordination within the CuO$_2$ chain, which is significantly larger than that of a Cu$^{+}$ ion (0.60 {\AA} \emph{CN}=2) within the CuO$_2$ dumbbell.\cite{Shannon1976}  Room temperature lattice parameters for LiCu$_{2-z}$Zn$_{z}$O$_{2}$ as a function of Zn substitution level is summarized in figure \ref{fig:fig04}.  We note a significant reduction of \emph{c} axis in concomitant with slight increase of \emph{a} and \emph{b} axes when crossing the Zn $\sim$ 5 $\%$ boundary, with minimal change on the cell volume.  These results suggest the success of Zn substitution to the Cu$^{2+}$ instead of the smaller Cu$^+$ site within the bridging O-Cu-O dumbbell, and the average lattice distortion must be intimately correlated to the occurrence of novel phase transitions found for \emph{z} $>$ 0.05 at low temperature to be discussed further in the text.

\subsection{Magnetic susceptibilities}

\begin{figure}
\begin{center}
\includegraphics[width=3.5in]{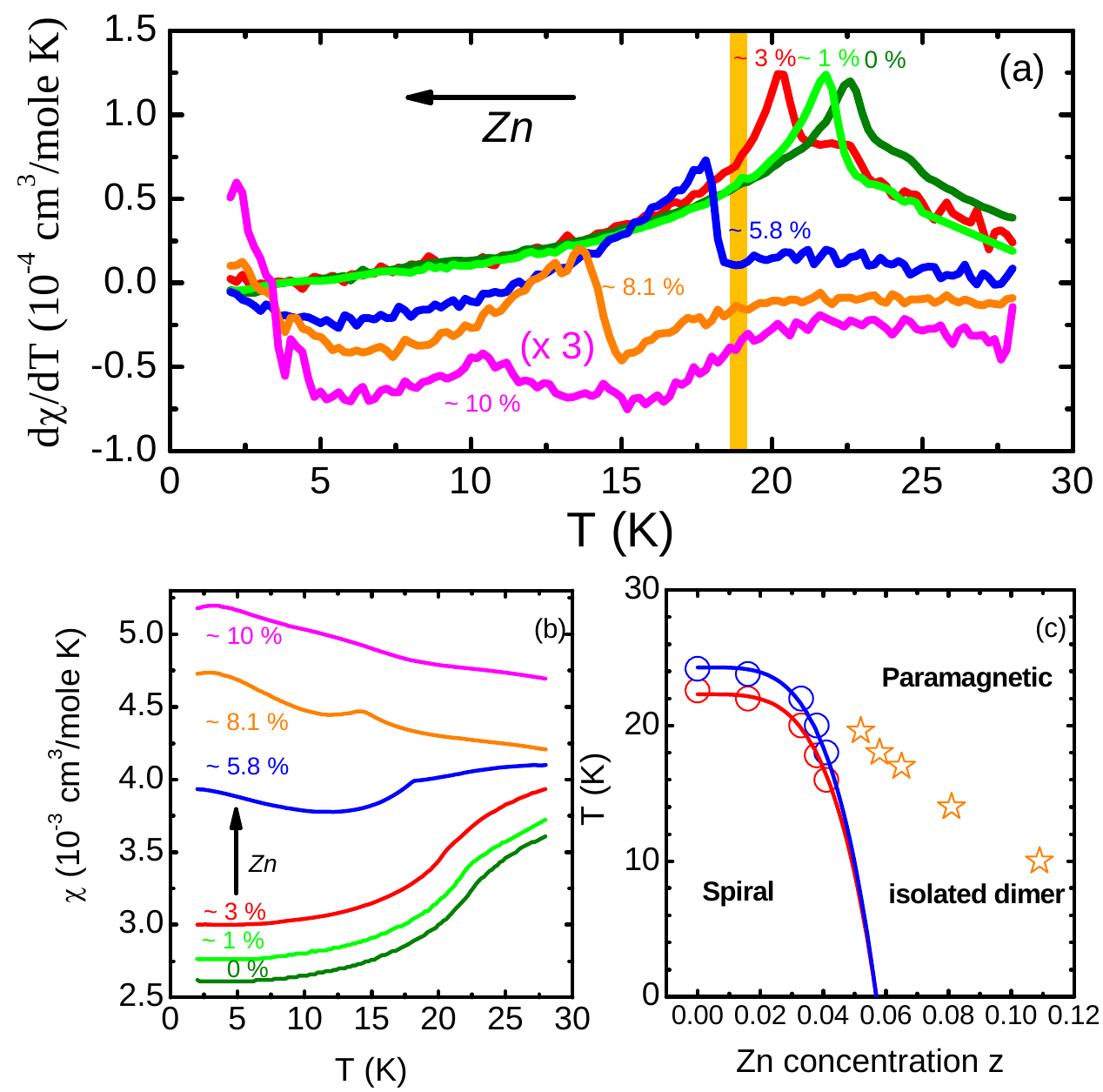}
\end{center}
\caption{\label{fig:fig05}(Color online) Temperature dependence of \emph{d$\chi$$_{ab}$/dT} for LiCu$_{2-z}$Zn$_{z}$O$_{2}$ along the $ab$ plane with \emph{z} = 0, 1.6, 3.4, 5.8, 8.1, and 10.9\%. For clarity, the signal for \emph{z} = 10.9\% is multiplied by three.  \emph{$\chi$(T)}$_{ab}$ data are shown in (b), which are normalized to the $\chi$(300K) value. The helical ordering temperature \emph{T$_h$} versus Zn content is summarized in (c).  Finite size scaling behavior is found for Zn below $\sim$ 5.5$\%$ for the two transition temperatures with \emph{H} applied along the \emph{ab} plane and \emph{c} direction.}
\end{figure}

\begin{figure}
\begin{center}
\includegraphics[width=3.5in]{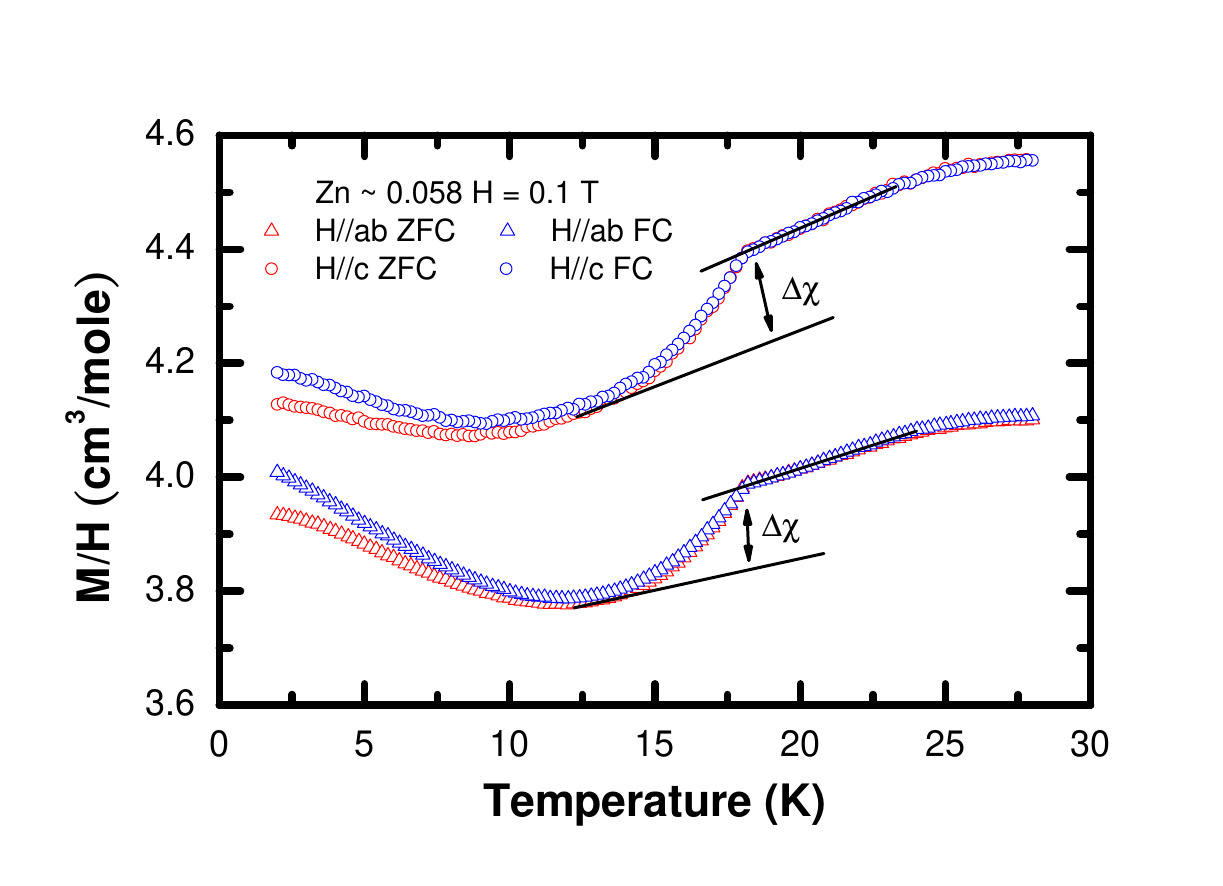}
\end{center}
\caption{\label{fig:fig06}(Color online) Magnetic susceptibilities for Zn $\sim$ 5.8 $\%$ sample with field applied along the \emph{ab} and \emph{c} directions. An estimate of susceptibility reduction is obtained from $\bigtriangleup$$\chi$/$\chi$ at \emph{T$_c$}.}
\end{figure}

%Chi(T) summary discussion
Figure \ref{fig:fig05} summarizes the complete evolution of Zn substitution effect shown in \emph{$\chi_{ab}$(T)} and \emph{d$\chi_{ab}$/dT(T)} data for all samples studied.  Low Zn substitution reveals typical signature of anisotropic helimagnetic ordering transition and reduces \emph{T$_h$} slightly below Zn $\sim$ 5.5 $\%$ per CuO$_2$ chain to show peaks of \emph{d$\chi$(T)/dT} at $\sim$ 20 and 22.5 K, which are slightly lower than that of the pristine crystal but similar to the nearly Li deficiency-free sample as reported previously.\cite{Seki2008, Hsu2008}   Across the Zn $\sim$ 5.5 $\%$ phase boundary, \emph{$\chi$(T)} transforms from a spiral spin ordering character toward a step-like crossover or a cusp shape near the phase transition as shown in figure \ref{fig:fig05}(b).  The different characters for the helical and the novel phase transition are displayed not only by the symmetry change of \emph{d$\chi$(T)/dT} peak shape but also by its smooth crossover ($\leq$ 5.5 $\%$) versus cusp shape ($\geq$ 5.5 $\%$) difference in \emph{$\chi$(T)}.

The lowering  \emph{T$_h$} for helical spin ordering with Zn $\leq$ 5.5 $\%$ per CuO$_2$ chain suggests that the helical spin ordering condition of $|$\emph{J$_2$/J$_1$}$|$ $>$ $\frac{1}{4}$ must be met still, which is not surprising considering the perturbed spin chains through low level nonmagnetic ion substitution.  Low Zn substitution does not break the original spiral spin ordering but only reduces its transition temperature slightly from $\sim$ 22 K to the lowest $\sim$ 20 K before two coexisting phases emerge.  It is interesting to note that the single phase sample of the lowest \emph{T$_h$} has been achieved through Li$_{0.99}$Cu$_2$O$_2$ before, and similar $\sim$ 20 K transition temperature is found when the Li deficiency is reduced to the minimum.\cite{Hsu2008}  Clearly the spiral spin ordering of LiCu$_2$O$_2$ system prefers an optimum level of Cu$^{2+}$ excess ($\sim$ 0.17/f.u.) instead of exactly equal amount of Cu$^+$ and Cu$^{2+}$, which must be strongly correlated to the required incommensurate periodicity of spiral spin ordering with modulation $\zeta$ = 0.174 along the chain \textbf{b}-direction, while the periodicity is intimately related to the \emph{J$_2$/J$_1$} ration classically.\cite{Huang2008}  Current phenomenon of \emph{T$_h$} reduction as a result of Zn substitution could be due to the finite chain effect, i.e., the Zn substitution cuts off the original helically ordered infinite chains before the optimal incommensurate modulation for the spiral ordering is broken.  While \emph{J$_1$} may be frustrated locally near the Zn site, the average \emph{J$_2$/J$_1$} ratio may not be modified very much at Zn levels lower than $\sim$ 5.5 $\%$, which in fact has been supported from our HTSE fitting results to be discussed below.

The Zn substitution above $\sim$ 5.5 $\%$ per CuO$_2$ chain introduces an intriguing phase transition which is hard to identify based on its \emph{$\chi$(T)} character of nearly isotropic cusp shape as shown in figure \ref{fig:fig06}.  Since the newly discovered phase transition is nearly isotropic for this quasi-two-dimensional system of specific spiral plane, it is unlikely that the transition to be antiferromagnetic ordering or even spin glass.  The step-like isotropic reduction of spin susceptibility implies the existence of a gapped phase that comes from part of the system, judging from its non-zero susceptibility at low temperature, even after a background subtraction of expected enhancing Curie contribution from Zn introduced isolated spins.  As shown in figure \ref{fig:fig06}, the ratio of $\chi$ step reduction versus peak $\chi$ value at \emph{T$_h$} ($\bigtriangleup$$\chi$/$\chi$ at \emph{T$_c$}) is very close to the Zn substitution level between $\sim$ 0.045 - 0.055, which implies that only part of the system falls to the gapped ground state and is proportional to the Zn level, i.e., high Zn substitution introduces a gapped phase that is proximate to the Zn centers.  On the other hand, more pronounced hysteresis is observed in particular below 5 K, which implies that glassy or irreversible domain effect occurs below \emph{T$_c$}.  Bobroff \textit{et al.} have proposed that Zn impurities could introduce collective spin freezing around each impurity center for spin gapped system at low temperature,\cite{Bobroff2009} which may also apply to the gapped phase in our system as suggested by the spin glass like behavior below $\sim$ 5 K.  Further muon spin resonance experiment is planned to examine this proposed explanation.

\begin{figure}
\begin{center}
\includegraphics[width=3.5in]{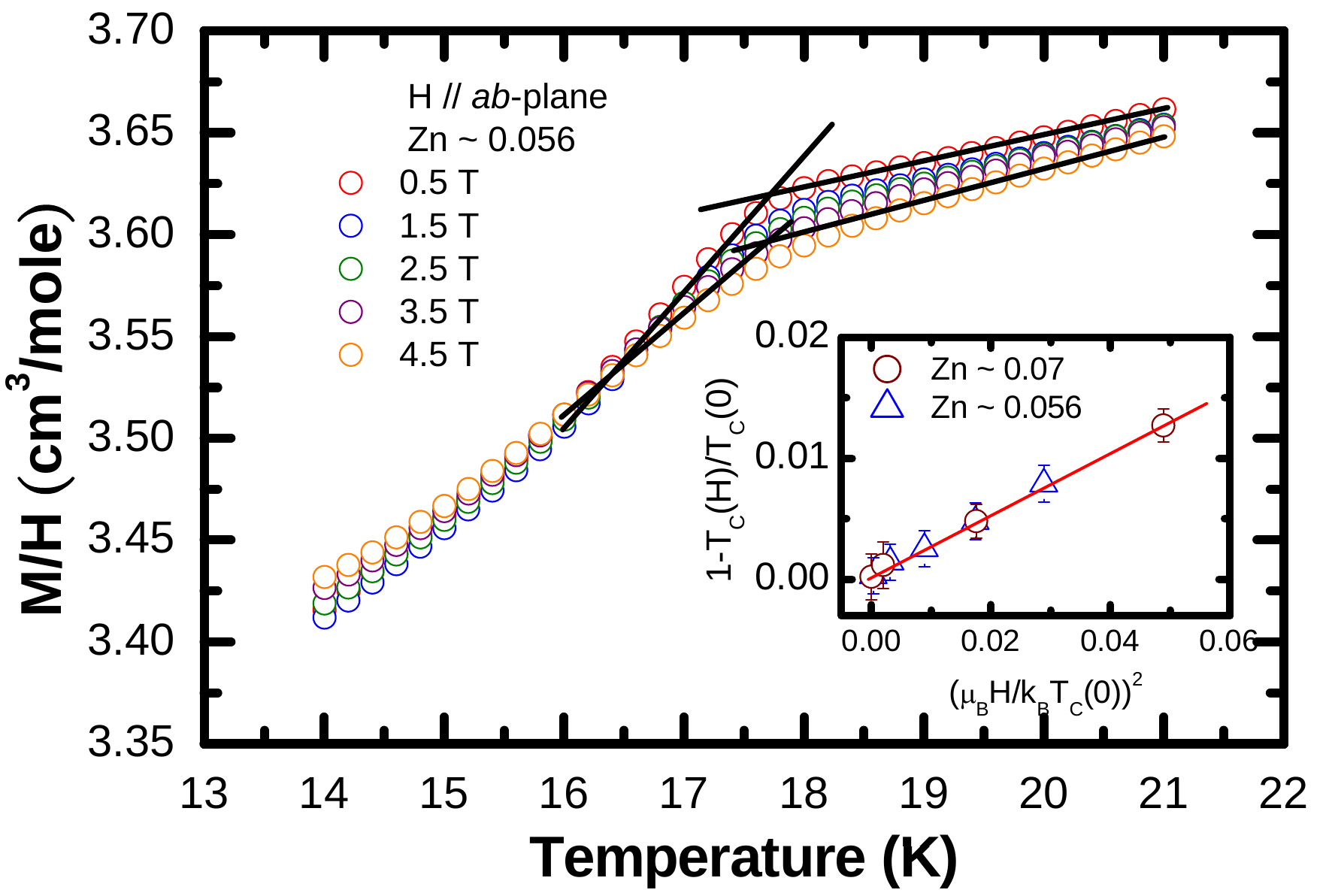}
\end{center}
\caption{\label{fig:fig07}(Color online) Magnetic susceptibilities of single crystalline LiCu$_{2-z}$Zn$_{z}$O$_{2}$ measured near the phase transition temperature \emph{T$_{c}$(H)} in the range of \emph{H} = 0.01 - 4.5 T.  \emph{T$_{c}$(H)} value is defined by the intersection of two linear fitted lines from above and below.  A universal plot of \emph{1-T$_{c}$(H)/T$_{c}$(0)} versus  [\emph{-$\mu_{B}$H/k$_B$T$_{c}$(0)}]$^2$, including two samples of different Zn levels above $\sim$ 5.5 $\%$, is shown in the inset with a linear fitting of slope $\sim$ 0.26.}
\end{figure}

%scaling discussion on field dependence
We traced magnetic field dependence of transition temperature of crystals with Zn level higher than $\sim$ 5.5 $\%$ and found the scaling of \emph{1-T$_{c}$(H)/T$_{c}$(0)} shows a convincing \emph{H$^2$} dependence as seen in the inset of figure \ref{fig:fig07}.  The transition temperatures have been determined by linear extrapolation from the upper and lower parts of the susceptibility data near \emph{T$_c$} before they deviate again due to Curie contribution, similar to that found in the Zn substituted spin Peierls compound CuGeO$_3$.\cite{Hase1993}  Bulaevskii \textit{et al.} have shown that commensurability effect in spin-Peierls (SP) transition would cause strong magnetic field dependence of \emph{T$_{c}$} in the limit of \emph{$\mu_B$H $\ll$ k$_{B}$T$_c$(0)}  to follow equation as\cite{Bulaevskii1978}

\begin{equation}
1-\frac{T_{c}(H)}{T_{c}(0)} \sim \beta [\frac{\mu_B H}{k_B T_{c}(0)}]^2.
\label{eq:two}
\end{equation}

\noindent Such field dependence is a result of band filling change that is sensitive to the competing commensurability-incommensurability requirement between spin and phonon for spin-Peierls transition.  We find that there is clear \emph{H$^2$} dependence for the reduced critical temperature.  Such \emph{H$^2$} field dependence has been observed in CuGeO$_3$ of well known spin-Peierls (SP) transition before.\cite{Hase1993}  The quadratic field dependence found in equation \ref{eq:two} also ruled out the possibility of ordinary structural transition due to phonon instability under strong field.\cite{Bulaevskii1978}  There is no evidence of structural transformation or signature of lattice doubling found based on our synchrotron x-ray structure analysis for the whole Zn substituted series (not shown).   The proportional constant $\beta$ shown in equation \ref{eq:two} is fitted to be $\sim$ 0.26, which is lower than the 0.46 found in the dimerized system CuGeO$_3$ but higher than the calculated value of 0.11 within Hartree-Fock approximation.\cite{Hase1993, Bulaevskii1978}  Moreover, we find such quadratic field dependence of \emph{T$_c$} is independent of Zn content as shown in figure \ref{fig:fig07}, which implies that the same universality class applies to the impact of Zn substitution above $\sim$ 5.5 $\%$; and this implication must be clarified further in the future both experimentally and theoretically.

\subsection{Finite size scaling}

%Finite size effect for Zn < 5% region discussion
Figure \ref{fig:fig05} shows that helical ordering transition temperature \emph{T$_h$(z)} for LiCu$_{2-z}$Zn$_z$O$_2$ decreases with higher \emph{z} before it drops precipitously for \emph{z} $\sim$ 0.055 critically.  The precipitous drop of \emph{T$_h$(z)} suggests a power law \emph{z} dependence of large exponent.  Fitting \emph{T$_h$(z)} with reduced temperature and \emph{z} in power law as

\begin{equation}
1-\frac{T_{h}(z)}{T_{h}(0)}=(\frac{z}{z_c})^n \propto L^{-\frac{1}{\nu}}
\label{eq:three}
\end{equation}

\noindent and we find the exponent $n$ to be near 4 as \emph{T$_h$(0)}, \emph{z$_c$} and \emph{n} are left as free parameters in the fitting.  An equally good fitting is obtained when $n$ is fixed at 4, and \emph{T$_h$(0)} and \emph{z$_c$} return to be 22.3(3) (for \emph{H}$\|$\emph{ab}), 24.3(3) (for \emph{H}$\|$\emph{c}) and 0.057(2), respectively.  The satisfactory fitting of equation~\ref{eq:three} suggests that finite size effect is able to describe the \emph{z} dependence of \emph{T$_h$}.  Finite size scaling has been applied in Monte Carlo simulation using limited size \emph{L} and its size extrapolated exponent $\nu$ to calculate the critical temperature in real infinite size matrix as described in equation~\ref{eq:three}.\cite{Fisher1972}  Since the critical exponent $\nu$ within mean field approximation to be 1/2, the fitted \emph{n}=4 suggests that \emph{T$_h$} reduction is a result of limited correlation length due to finite size, i.e., \emph{L(z)} $\propto$ \emph{1/z$^2$}.  The most natural model that can explain such \emph{z} dependence of \emph{L(z)} would be a picture of two-dimensional helical ordered domains which are cut off by defect or secondary phases formed domain boundaries, the higher the Zn level the smaller the domain size $\sim$ \textbf{a}(\textbf{b})/\emph{z$^2$}.  Assuming dopants are distributed at the two-dimensional domain boundaries, the effect of lower \emph{T$_h$} can be explained as a result of cut-off correlation length by the domain boundaries.  Since the in-plane magnetic correlation length $\xi_{ab}$ for the pure LiCu$_2$O$_2$ has been estimated to be $\sim$ 1200 {\AA},\cite{Huang2008} which is very close to the domain size estimated for \emph{z} $\sim$ 0.055 at the phase boundary, i.e. \emph{L} = \emph{b}/0.055$^2$ $\sim$ 1000 {\AA}.  It is implied that long range helimagnetic spin ordering is destroyed totally once the domain size becomes smaller than the helical magnetic correlation length.

The original undoped sample has a quasi-two-dimensional matrix of helical ordered incommensurate spin structure with periodicity about 5.7$\bf{b}$, i.e., modulation vector $\zeta$ $\sim$ 0.174 along the \textbf{b}-axis with pitch angle 62.6$^\circ$.\cite{Huang2008}   We find that the phase boundary for Zn $\sim$ 5.5$\%$ is very close to the onset of non-zero probability of aggregated (more than one) Zn per 6$\textbf{b}$$\times$2\textbf{a} magnetic supercell of incommensurate helical ordering, i.e.,  starting from 2 out of $\sim$ 35 Cu sites are substituted by Zn.  The estimated phase boundary agrees very well with the phase boundary \emph{z$_c$} obtained from the finite-size scaling fitting as discussed previously.  The phase boundary obtained from simple probability analysis suggests that long-range helical ordering is suppressed only when more than one Zn exist per unit superlattice of incommensurate helical ordering.  For Zn substituted sample prepared from high temperature melt, it is reasonable to assume that Zn ions distribute randomly.  Although it is unreasonable to assume Zn to be mobile at low temperatures, we suspect that the comparable and competing (frustrating) couplings \emph{J$_1$}, \emph{J$_2$} and \emph{J$_\bot$} could transport spin degree of freedom from Zn centers to the domain boundaries that surround helically ordered spins.  Such phenomenon could be viewed as a  microscopic phase separation with localized spins confined within domain boundaries and encircle domains with spins of persisted helical ordering, an interesting contrast to the phase separation resulting from one-dimensional charge stripe separated antiferromagnetic domains in La$_{2-x}$Sr$_x$CuO$_4$.\cite{Cho1993}

\begin{figure}
\begin{center}
\includegraphics[width=3.5in]{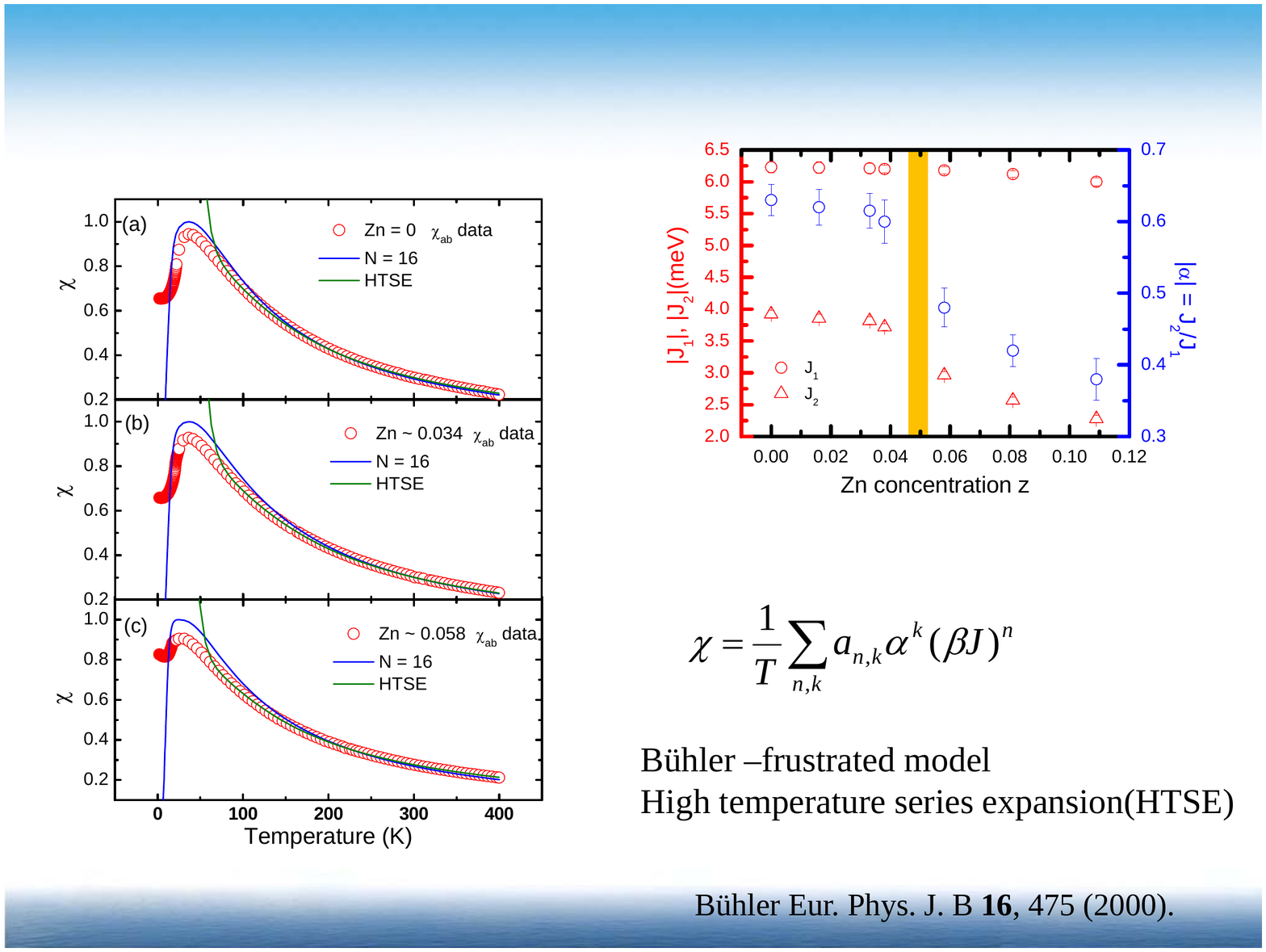}
\end{center}
\caption{\label{fig:fig08}(Color online) High temperature series expansion (HTSE) fitting and exact Hamiltonian diagnolization using 16 spin ring simulation results of magnetic susceptibility data for samples with Zn $\sim$ 0, 3.4 and 5.8 $\%$.}
\end{figure}

\begin{table}
\begin{center}
\caption{\label{tab:tableII} Exchange coupling constants \emph{J$_{1}$} and $\alpha$=\emph{J$_{2}$/J$_{1}$} obtained from high temperature series expansion (HTSE) method and exact diagonalization of the Hamiltonian with 16 spin ring (N=16) simulation.  Estimations from neutron scattering spin wave analysis and LDA calculation are listed for comparison.}
\begin{tabular}{cccccccccc}
 \hline
 \hline
Zn content & \multicolumn{2}{c}{HTSE} & N=16  & \multicolumn{3}{c}{Neutron study\cite{Masuda2005}} &\multicolumn{3}{c}{LDA\cite{Gippius2004}}\\
(EPMA) & J$_{1}$(meV) & $\alpha$ & $\alpha$ & J$_{1}$(meV) & $\alpha$ & J$_{\bot}$ & J$_{1}$(meV) & $\alpha$ & J$_{\bot}$ \\
 \hline
 \hline
0 & -6.23(1) & -0.63(2) & -0.578 & -5.95$^{\dag}$ & -0.62$^{\dag}$  & 0.9$^{\dag}$   & -8  & -1.777 & 5.7 \\
    &          &          &        & -7$^{\ddag}$    & -0.535$^{\ddag}$ & 3.4$^{\ddag}$   &     &        &     \\
0.016(3) & -6.22(2) & -0.62(1) &        &                       &                        &                       &     &        &     \\
0.034(2) & -6.21(3) & -0.61(1) & -0.54  &                       &                        &                       &     &        &     \\
0.038(2) & -6.20(2) & -0.60(3) &        &                       &                        &                       &     &        &     \\
0.058(2) & -6.18(2) & -0.48(4) & -0.468 &                       &                        &                       &     &        &     \\
0.081(3) & -6.12(3) & -0.42(2) &        &                       &                        &                       &     &        &     \\
0.109(3) & -6.00(1) & -0.38(1) &        &                       &                        &                       &     &        &     \\
 \hline
 \hline
\end{tabular}
\end{center}
$\dag$ {model 1 of reference \cite{Masuda2005}}\\
$\ddag$ {model 3 of reference \cite{Masuda2005}}
\end{table}

\begin{figure}
\begin{center}
\includegraphics[width=3.5in]{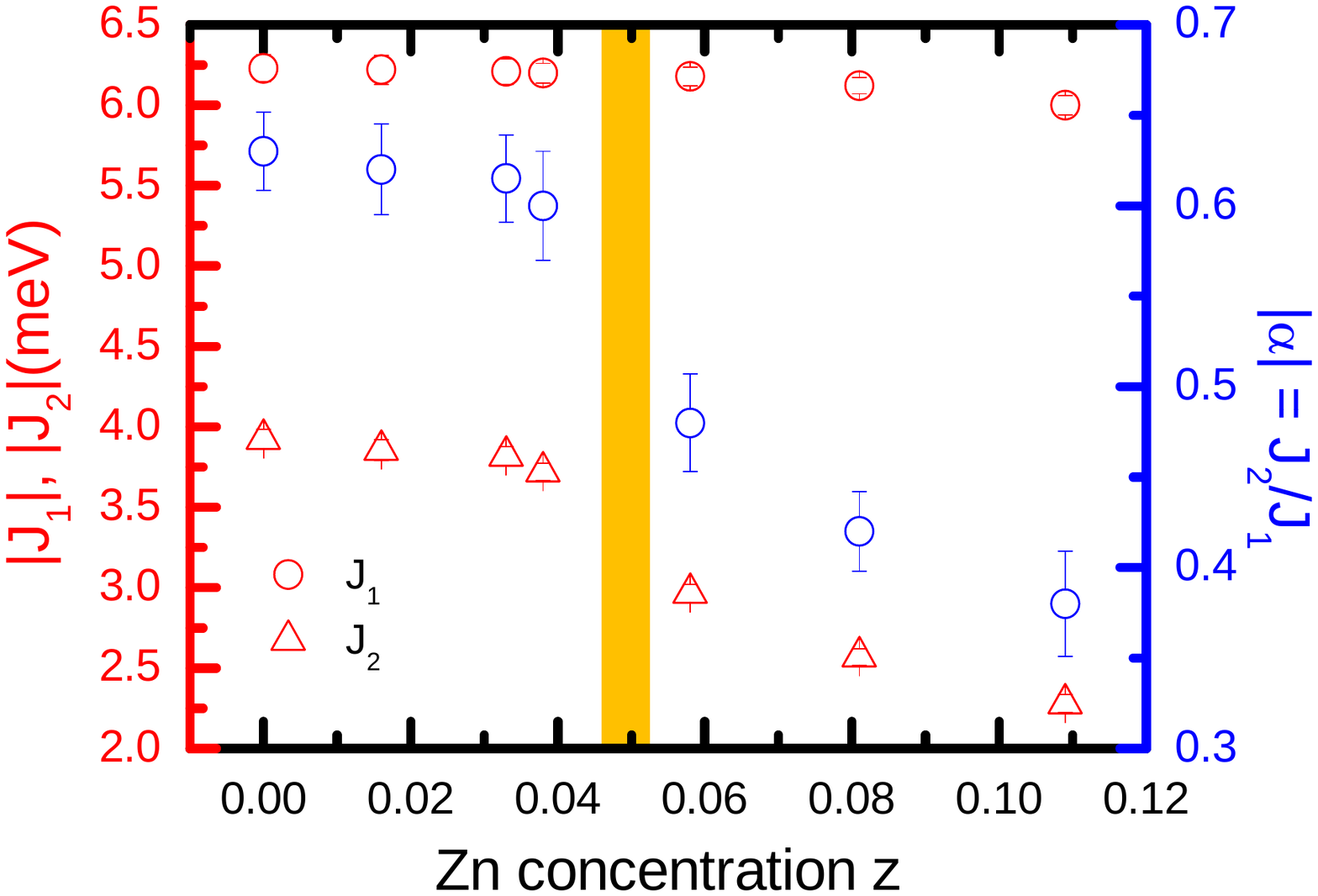}
\end{center}
\caption{\label{fig:fig09}(Color online) \emph{J$_1$}, $\alpha$ and \emph{J$_2$} (converted from $\alpha$) obtained from HTSE fitting results versus Zn substitution level for LiCu$_{2-z}$Zn$_z$O$_2$.}
\end{figure}

\subsection{HTSE and exact diagonalization fittings}

% fitting of HTSE and N=16
In order to trace the frustrating couplings \emph{J$_1$} and \emph{J$_2$} ($\alpha$=\emph{J$_2$/J$_1$}) for this helimagnetic system, we applied high temperature series expansion (HTSE) and exact Hamiltonian diagonalization methods to fit the magnetic susceptibility data for the entire Zn substitution range.  Starting from Hamiltonian written as equation \ref{eq:one}, the HTSE uses expansion coefficients for the frustrated quantum spin chain model calculated by B\"{u}hler $\it{et~al.}$\cite{Buhler2000}  Exact diagonalization of the Hamiltonian using 16 spin ring (N=16) calculation is used to provide satisfactory temperature dependent fitting at high temperature range after peak normalization and background subtraction.  Both HTSE and N=16  fittings return satisfactory results on high temperature range and agree very well on the estimation of \emph{J$_1$} and $\alpha$ as shown in figure \ref{fig:fig08} and summarized in figure \ref{fig:fig09} and table \ref{tab:tableII}.  These results are in agreement with those obtained from LDA calculation and neutron scattering spin wave analysis.\cite{Gippius2004, Masuda2005}  The fitting results using N=16 finite chain are in agreement at high temperature range with HTSE within 6(1) $\%$, although the low temperature region deviates significantly as a result of gapped finite size chain approximation.  We find \emph{J$_1$} decreases smoothly, crossing the $\sim$ 5 $\%$ Zn phase boundary by $\sim$ 4 $\%$ only, while \emph{J$_2$} (calculated from \emph{J$_1$} and $\alpha$) reduction is far more pronounced.  The significant \emph{J$_2$} reduction through Zn substitution moves this frustrated system closer to the quantum critical point near $\alpha$ = -1/4 as shown in figure \ref{fig:fig03}, and complex periodicity may have occurred.\cite{Bursill1995, Schmitt2009}

\begin{figure}
\begin{center}
\includegraphics[width=3.5in]{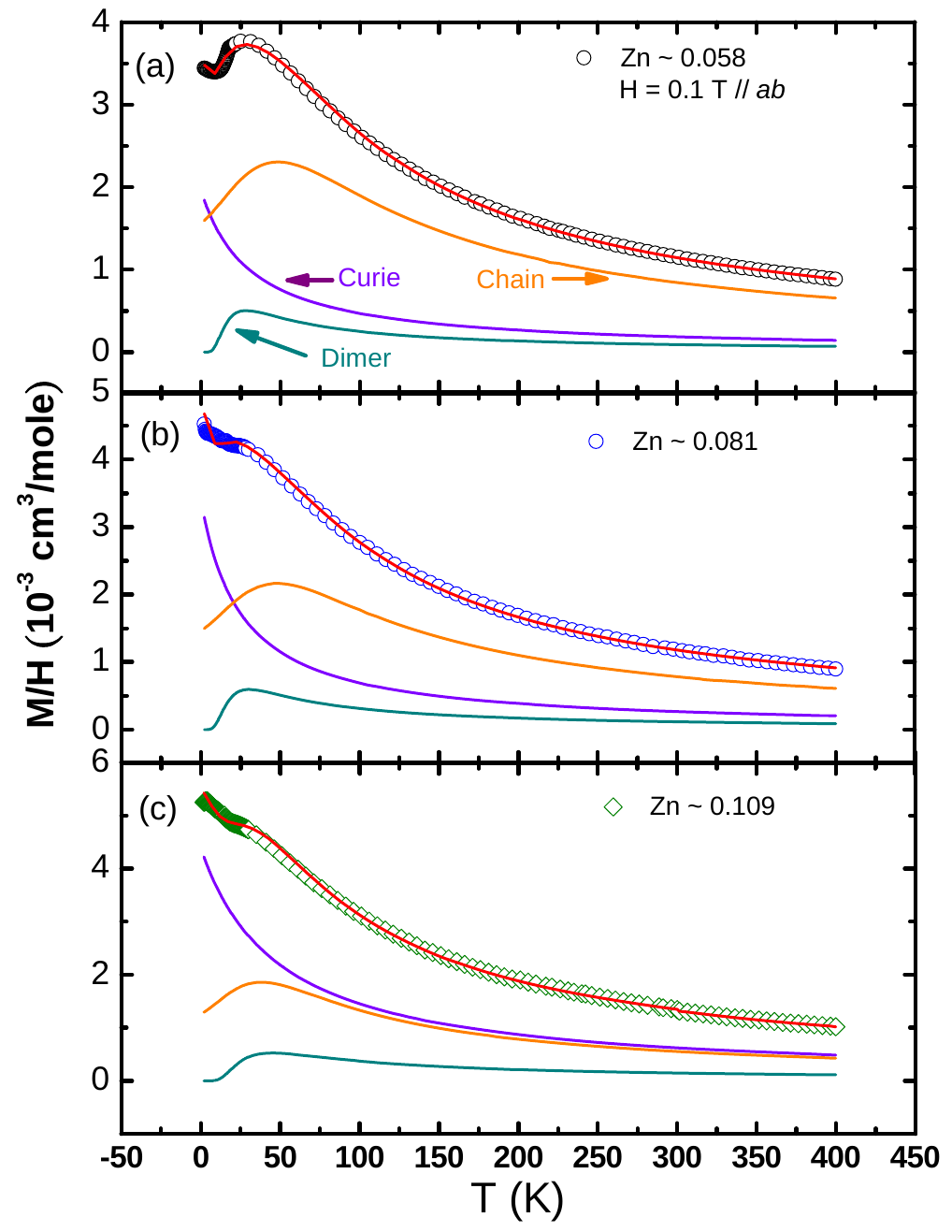}
\end{center}
\caption{\label{fig:fig10}(Color online) Combined HTSE, isolated dimer and Curie-Weiss model fitting.}
\end{figure}

%Fitting of HTSE+Dimer+Curie

While high Zn substitution is proposed to introduce isolated dimers into the original long-range helical spin ordered system, the magnetic susceptibility measurement results should reflect such assumption, i.e.,  the final form of \emph{$\chi$(T)} for Zn $\geq$ 5.5 $\%$ must include contributions from a gapped state of isolated dimers as well as the isolated spins that shows paramagnetic behavior.  We fit the high Zn data additionally with combined HTSE, isolated dimer and Curie contributions as shown in figure \ref{fig:fig10}, following equation

\begin{eqnarray*}
\chi &=  \chi_{_{0}} + N_{chain} \chi_{_{chain}} + N_{_{D}} \chi_{_{Dimer}} + N_{_{CW}} \chi_{_{CW}}, \\
     &=  \chi_{_{0}} + N_{chain} \frac{1}{T} \sum_{n,k} a_{n,k} \alpha^k (\frac{J_{_{1}}}{k_{_{B}}T})^n + N_{D}\frac{N_A g^2 \mu_{_B}^2}{k_{_{B}}T (3 + e^{\frac{J_{_{D}}}{k_{_{B}}T}})} + N_{_{CW}} \frac{C}{T-\Theta}. (4)
\label{eq:Buhler}
\end{eqnarray*}

\noindent The only constraint used in the fitting process is the normalization of \emph{N$_{chain}$+N$_{dimer}$+N$_{Curie}$}=1.  The complete fitted results are summarized in table \ref{tab:tableIV}.  Temperature dependence of isolated dimers with spin gap size \emph{J$_D$} follows the Bleany-Bowers equation, which can be reduced to describe the low temperature spin gap behavior and the Curie-Weiss behavior for dimers at high temperature.\cite{Johnson1984}  The fitted results using equation~\ref{eq:Buhler} are not perfect, possibly due to the incomplete HTSE chain description, neglected interchain coupling, and the possible short-range coupling among dimers; however, it does provide a better fitting below \emph{T$_c$} for higher Zn above $\sim$ 5.5 $\%$.  The amount of dimer phase \emph{N$_D$} is nearly proportional to the Zn content level, which supports a picture of Zn introduced isolated dimers favorably.  Moreover, the fitted values of \emph{J$_D$} agree very well with the LDA and neutron scattering estimation of \emph{J$_\bot$} (see table \ref{tab:tableII}), which supports a scenario of dimer formation as a result of O-Cu$^+$-O bridged nontrivial interchain coupling \emph{J$_\bot$}.  The Curie contribution from isolated spins, \emph{N$_{CW}$}, increases drastically for \emph{z} $\sim$ 0.11, which is reflected on the weaker and broadened phase transition as shown in figure \ref{fig:fig05}(a) also.  Following the same phase separation model suggested here, we find that the magnetic susceptibility reduction at \emph{T$_c$} ($\Delta\chi$/$\chi$(\emph{T$_c$})) also reflects the spin gapped phase proportion as summarized in table \ref{tab:tableIV}, before it is smeared by the significantly enhanced Curie contribution at \emph{z} $\sim$ 0.11.  Both the \emph{$\chi$(T)} model fitting and the susceptibility reduction at \emph{T$_c$} have demonstrated satisfactory agreement on the fraction of isolated dimer phase and the Zn content for \emph{z} $\geq$ 5.5 $\%$, which supports the interpretation of the special magnetic susceptibility anomalies near \emph{T$_c$}  (see figure \ref{fig:fig05}) based on a scenario of phase separation.

\begin{table}
\begin{center}
\caption{\label{tab:tableIV}Magnetic susceptibility fitting results of high Zn doping level by the proposed model of combined spin chain, isolated dimers and spins following equation.~\ref{eq:Buhler}.}
\begin{tabular}{cccc}
 \hline\hline
 & \emph{z}=0.06(1) & \emph{z}=0.08(1) & \emph{z}=0.11(1) \\
 \hline
N$_{chain}$                         & 0.82     & 0.76     & 0.52      \\
J$_{c}$  (K)                        & 37.5(2)  & 37.0(3)  & 29.5(5)   \\
 \hline
N$_{_{CW}}$                         & 0.1      & 0.14     & 0.35      \\
$\Theta$ (K)                        & -31.6(9) & -25.6(4) & -49.4(7)  \\
C                                   & 0.62(6)  & 0.62(4)  & 0.60(2)   \\
 \hline
N$_{D}$                             & 0.08     & 0.10     & 0.13      \\
J$_{D}$ (K)                         & 45.8(6)  & 48.7(5)  & 91.8(3)   \\
 \hline\hline
$\Delta\chi$/$\chi$(\emph{T$_c$}) (\%)           & 4.5(1)   & 7(1)     & 0.5(3)    \\
 \hline\hline
\end{tabular}
\end{center}
\end{table}

\subsection{NMR results}

%NMR data discussion

\begin{figure}
\centering
\includegraphics[width=3.5in]{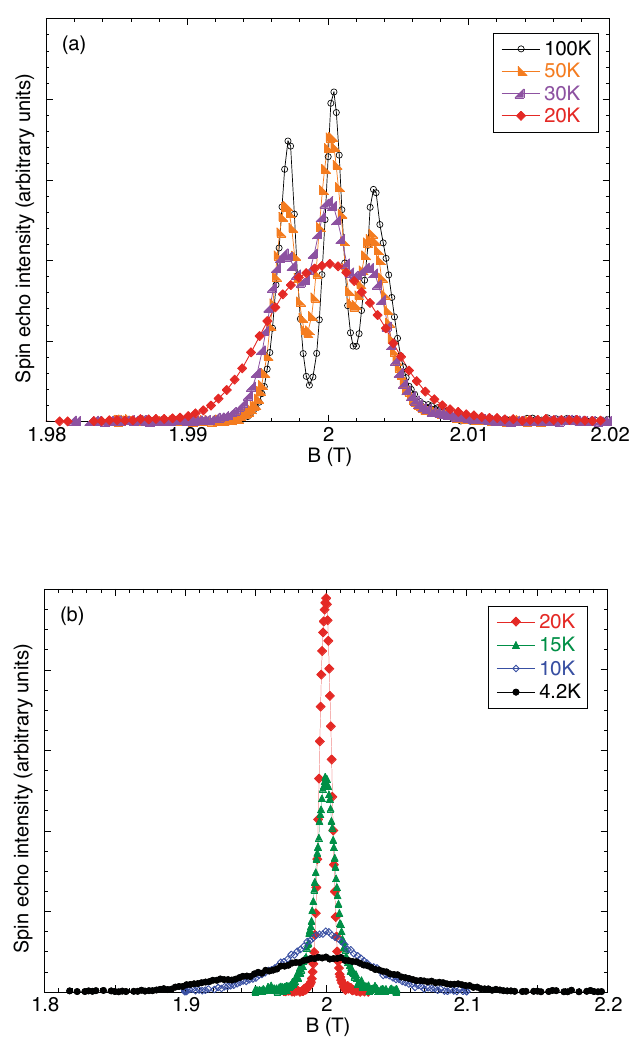} \\
\caption{\label{fig:fig11} (Color Online) Representative results of field-swept  $^{7}$Li NMR lineshapes measured at  $f = 33.095$~MHz, (a) at 20~K or above, and (b) at 20~K or below.  The external magnetic field was applied along the crystal \emph{c}-axis.  Notice that the scales of horizontal axis are different by a factor of ten between the two panels.}
\end{figure}

To gain further insight on the nature of the anomaly observed in the temperature dependence of $\chi$ for heavily Zn-doped crystals, we also carried out NMR measurements for $^{7}$Li  (nuclear spin $I=3/2$, nuclear gyromagnetic ratio $^{7}\gamma_{n}/2\pi = 16.546$~MHz/Tesla) in LiCu$_{2-z}$Zn$_z$O$_2$ (\emph{z} $\sim$ 0.058) by applying an external magnetic field $B$ along the crystal \emph{c}-axis.  We chose the NMR carrier frequency $f = 33.095$~MHz so that we could compare our NMR results for Zn-doped sample with an earlier NMR report for undoped LiCu$_2$O$_2$ by Gippius \textit{et al.} \cite{Gippius2004}  At 100~K, we observe a typical NMR lineshape expected for nuclei with spin $I=3/2$ with three distinct peaks split by the \emph{c}-axis component of the nuclear quadrupole interaction, $\nu_{Q}^{c}$.   As shown in figure \ref{fig:fig11}, the central peak near $B = 2$~Tesla arises from the nuclear spin $I_{z}=+1/2$ to $-1/2$ transition, while two additional peaks near  $B=1.997$~Tesla and 2.003 are from the $I_{z}=\pm3/2$ to $\pm1/2$ transitions.  From the splitting, 0.0030~Tesla, between the peaks, we estimate $\nu_{Q}^{c} = ^{7}\gamma_{n} \cdot 0.0030 = 0.05$~MHz.

\begin{figure}
\centering
\includegraphics[width=3.5in]{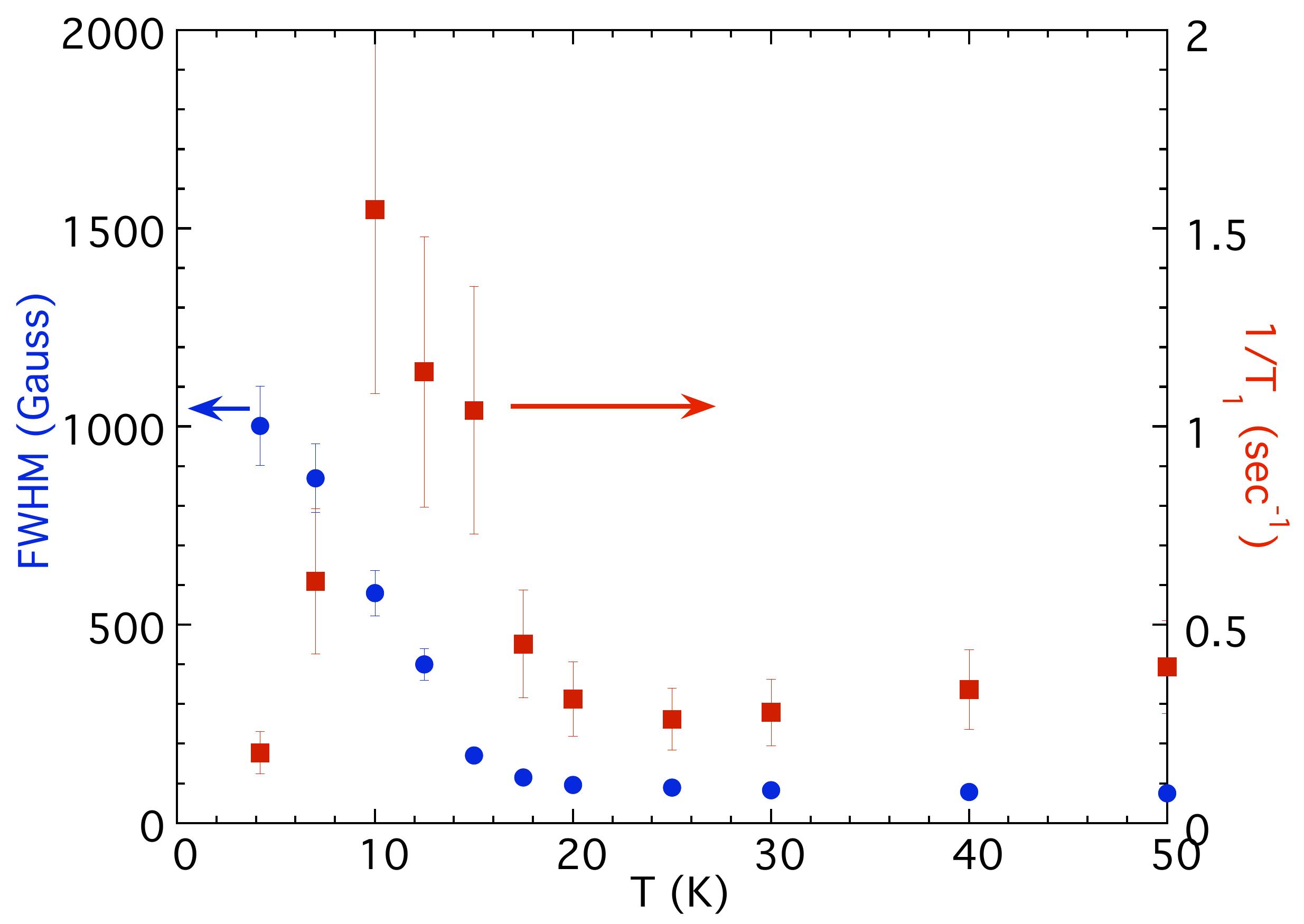}
\caption{\label{fig:fig12} (Color Online) The temperature dependence of (a) the FWHM of the $^{7}$Li NMR lineshape, and (b) the nuclear spin-lattice relaxation rate $1/T_{1}$ measured at the center of the lineshape.}
\end{figure}

Upon further cooling, all three peaks progressively become broader.  In principle, there are two possible mechanisms of the NMR line broadening in a paramagnetic material, i.e., magnetic effects or quadrupolar effects.  In the first scenario, all three peaks should broaden by an equal amount, while only the satellite peaks broaden in the second scenario.  The observed lineshapes at 50~K and 30~K clearly show that all three peaks are equally broadened, hence we conclude that there exists slight magnetic inhomogeneity.  This is consistent with the fact that Zn$^{2+}$ ions substituted Cu$^{2+}$ sites to remove the electron spin $S=1/2$, thus introducing random disorder in the matrix of $S=1/2$.   The splitting $\nu_{Q}^{c}$ between the peaks remains unchanged within experimental uncertainties down at least to 30~K.  This implies that EFG (Electric Field Gradient) at the $^{7}$Li site does not change significantly in this temperature range.  Below 20~K, the line broadening becomes so severe that we can no longer observe the quadrupole splitting between the three peaks.  At 10~K and 4.2~K, we observe an extremely broad lineshape with a very broad peak centered around 2~Tesla and two equally spaced shoulders near 1.92~Tesla and 2.08~Tesla.   We summarize the temperature dependence of the FWHM (Full Width at Half Maximum) of the NMR lineshape in figure \ref{fig:fig12}.  The FWHM begins to increase dramatically below $\sim18$~K, where $\chi$ in figure \ref{fig:fig06} shows a well-defined kink.

It is very instructive to compare these NMR lineshapes with those observed for undoped LiCu$_2$O$_2$ by Gippius \textit{et al.} \cite{Gippius2004} under nearly identical experimental conditions.  Our results above 30~K are very similar to the undoped case, but our lineshape at 4.2~K is strikingly different from that observed for LiCu$_2$O$_2$.  In the latter case, Gippius \textit{et al.} demonstrated that two large "horns" develop at 1.95 and 2.05~Tesla, accompanied by two additional horns near 1.9 and 2.1~Tesla (see the upper panel of figure 3 by Gippius \textit{et al.}).  They also demonstrated that these four horns are the consequence of  the distribution of hyperfine magnetic fields at $^{7}$Li sites, arising from an incommensurate spiral modulation of magnetic moments.  Our NMR lineshape at 4.2~K does not exhibit such sharp horn structures, but the overall shapes do bear a strong resemblance to each other.  The smearing of the sharp horns in our case is naturally understood as the consequence of disorder induced by Zn ions in the incommensurate modulation of magnetic moments.

The establishment of this low temperature modulated state with strong disorder is unlikely to be due to a second order phase transition.  In the case of a conventional second order magnetic phase transition, the nuclear spin-lattice relaxation rate $1/T_{1}$ would diverge at the phase transition temperature because of the critical slowing down of the magnetic moments.  In contrast, $1/T_{1}$  measured at the center of the $^{7}$Li NMR lineshape does not diverge near 18~K in the present case.   Instead, we observe a broad hump centered around 12~K.  This is consistent with glassy development of the modulated ground state.  Such glassy behavior has also been demonstrated by the pronounced thermal hysteresis in the magnetic susceptibility measurement shown in figure \ref{fig:fig06}.  It is reasonable to assume that the NMR results describe properties of the major spin population in the system, besides the isolated dimers or spins suggested above.  The glassy development below $\sim$ 12 K could be coming from the frozen short-range ordered spins within the frustrated finite spin chains.

\begin{figure}
\begin{center}
\includegraphics[width=3.5in]{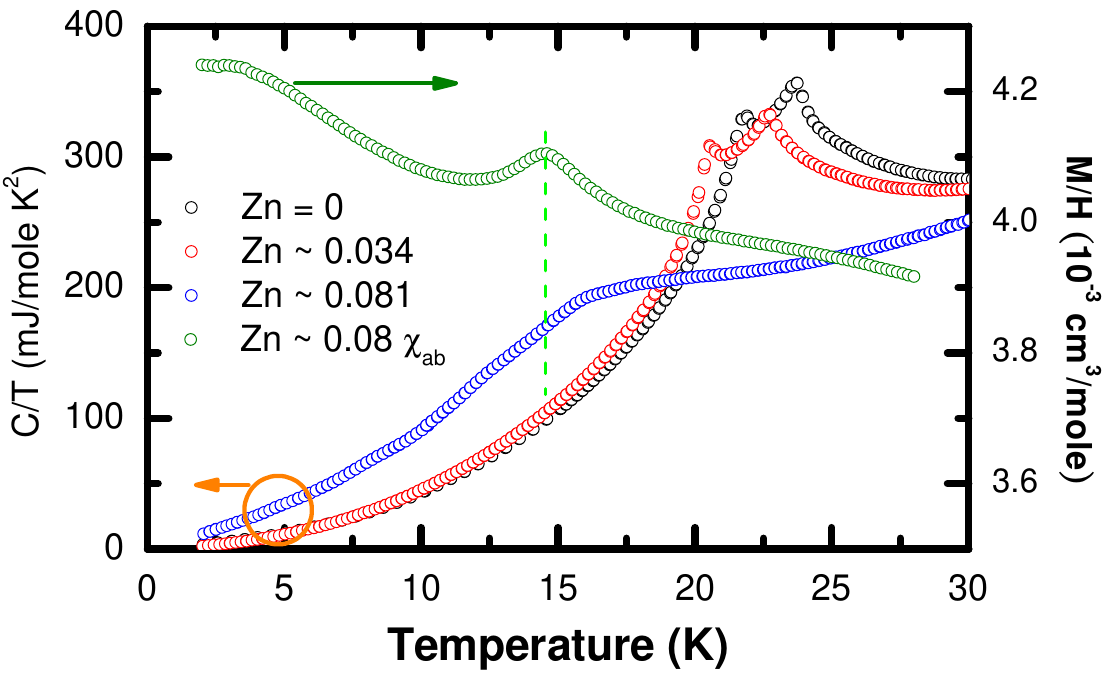}
\end{center}
\caption{\label{fig:fig13}(Color online) Specific heat of LiCu$_{2-z}$Zn$_{z}$O$_{2}$ with Zn = 0, 0.034, and 0.081. $\chi_{ab}$ for Zn = 0.081 is displayed also to show \emph{T$_c$} defined by the cusp shape anomaly.  }
\end{figure}

\subsection{\label{sec:level2}Specific heat\protect\\}

%specific heat discussion
Specific heat for samples with different levels of Zn substitution are shown in figure \ref{fig:fig13}.   The undoped and low Zn ($\leq$ 5.5 $\%$) samples show typical double peaks which correspond to the two distinct transition temperatures along the \textbf{ab}- and \textbf{c}-directions respectively for the spiral spin ordering.  Low Zn sample of 3 $\%$ substitution does not alter the transition width and simply shift the transition doublet to slightly lower, which indicates the robust nature of the helical spin ordering on zero spin perturbation.  On the other hand, a significantly broader specific heat anomaly is observed near $\sim$ 14 K for Zn $\sim$ 8 $\%$ sample, in fact there are two very broad peaks sitting above and below \emph{T$_c$} as defined by the cusp of \emph{$\chi$(T)}.  These broad anomalies suggest that either the phase transition is of short range in nature, or significant inhomogeneity exists in this Zn substituted crystal.  The existence of broad \emph{C$_{p}$} peaks together with the clear \emph{$\chi$(T)} cusp observed at \emph{T$_c$} suggests that short-range ordering of spin glass-like behavior occurs.  Although it is tempting to interpret the coexisting broad \emph{C$_p$} anomaly and $\chi$(\emph{T$_c$}) cusp to be coming from a well-defined spin glass phase transition, we cannot ignore the fact that there are two broad peaks of \emph{C$_p$} across \emph{T$_c$}, besides, the thermal hysteresis of \emph{$\chi$(T)} is not pronounce enough immediately below \emph{T$_c$} and only developed at much lower temperature as suggested by both the 1/\emph{T$_1$} and \emph{$\chi$(T)} below $\sim$ 5-10 K.  A reasonable interpretation can be obtained following our proposed phase separation scenario, i.e., we might view the broad "doublet" to be coming from short-range ordering of the frustrated spins (of reduced \emph{J$_2$/J$_1$} ratio) on the finite spin chain, which has been cut short by the Zn ions, while the cusp at $\chi$(\emph{T$_c$}) is a result of spin gap opening from those isolated spin dimers.

Magnetic entropy \emph{$\Delta$S$_M$} can be extrapolated from \emph{C$_p$} after the subtraction of lattice contribution and integrating from 2 to 160 K.  The entropy difference between the pure and the $\sim$ 8$\%$ Zn substituted samples is estimated to be $\sim$ 580 mJ/mole K, which is rather close to a portion (\emph{z}) of the total Cu$^{2+}$ spins in the system as \emph{zR*ln(2S+1)}=0.081*8.3*ln2=470 mJ/mol K, i.e., due to the missing spins for gapped singlet formation as a result of Zn substitution.  It is puzzling to note that the integrated \emph{$\Delta$S$_M$$\sim$} 18 J/mole K is almost three times larger than the theoretical expectation, i.e., \emph{$\Delta$S$_M$} = \emph{R}ln 2 = 5.8 J /mole K for a \emph{S}=1/2 system. This discrepancy must be due to certain factors other than phonon or spin. A possible candidate could be the long- or short-range ferroelectric ordering.

\begin{figure}
\begin{center}
\includegraphics[width=3.5in]{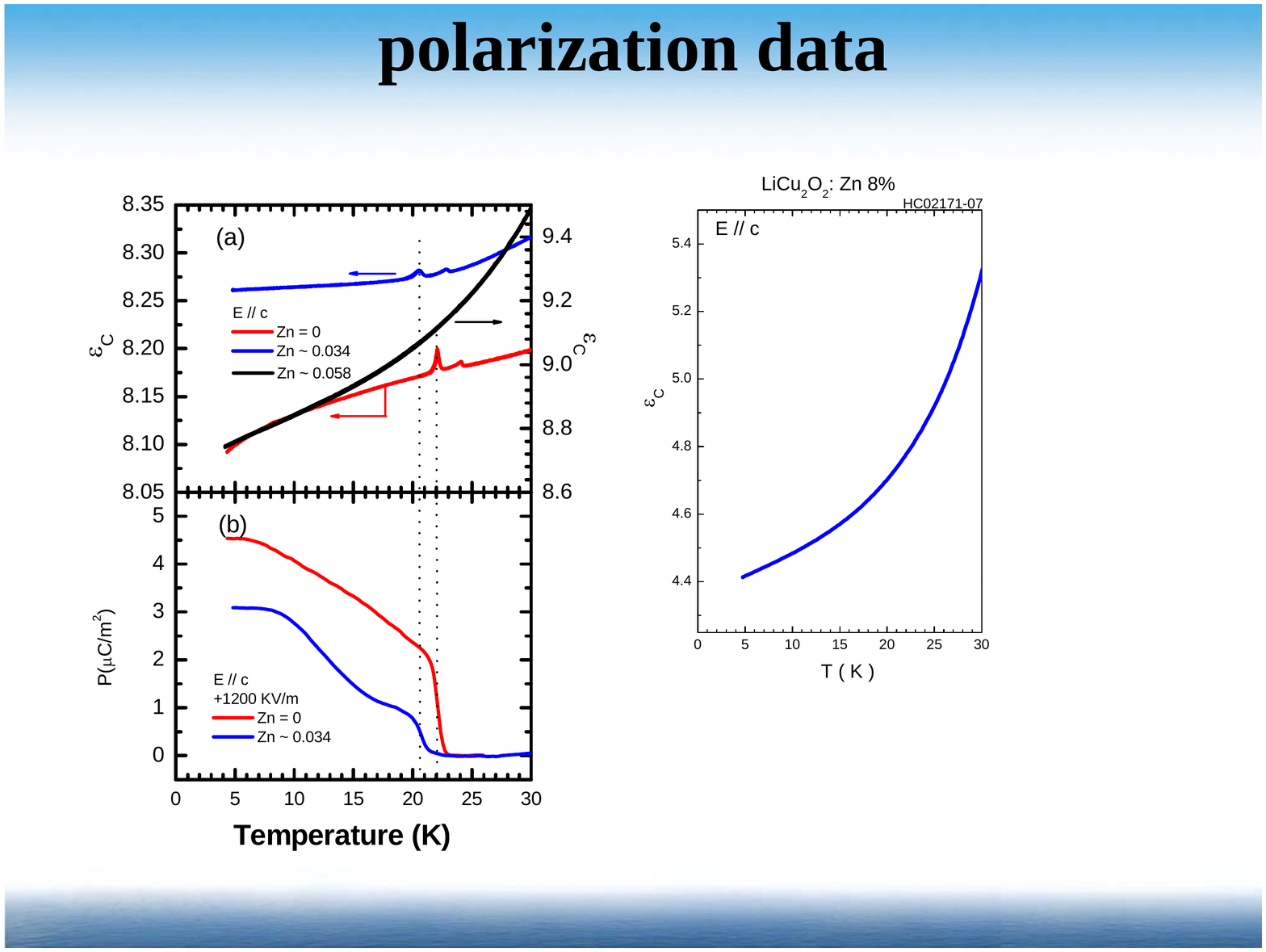}
\end{center}
\caption{\label{fig:fig14}(Color online) Electric polarization and dielectric response along the \emph{c}-direction for LiCu$_{2-z}$Zn$_{z}$O$_{2}$ with Zn = 0, 0.034, and 0.058.  An electric field of 1200 kV/m has been applied above the transition temperature before each measurement. }
\end{figure}

\subsection{Electric polarization}

%electric polarization
Dielectric constant and electric polarization have been measured for LiCu$_{2-z}$Zn$_z$O$_2$ (\emph{z} $\sim$ 0, 0.034, 0.058) single crystals along the \emph{c}-direction and shown in figure \ref{fig:fig14}.  The undoped sample shows the onset of electric polarization below $\sim$ 22 K and two anomalies near 22 and 24 K in dielectric constant similar to those reported in the literature.\cite{Seki2008}  3.4$\%$ Zn substitution reduces the spiral ordering temperature and electric polarization slightly.  5.8$\%$ Zn substitution sample shows no trace amount of electric polarization near its magnetic transition temperature near 14 K as indicated in figure \ref{fig:fig05}.  Spin-current or inverse Dzyaloshinskii-Moriya mechanism has been applied to interpret the origin of electric polarization through spiral spin ordering and has been tested by chirality reversion experiment.\cite{Katsura2005, Seki2008}  Current dielectric constant measurement results once again confirm that long-range spiral spin ordering is a prerequisite for the spontaneous electric polarization in current system.

\begin{figure}
\begin{center}
\includegraphics[width=3.5in]{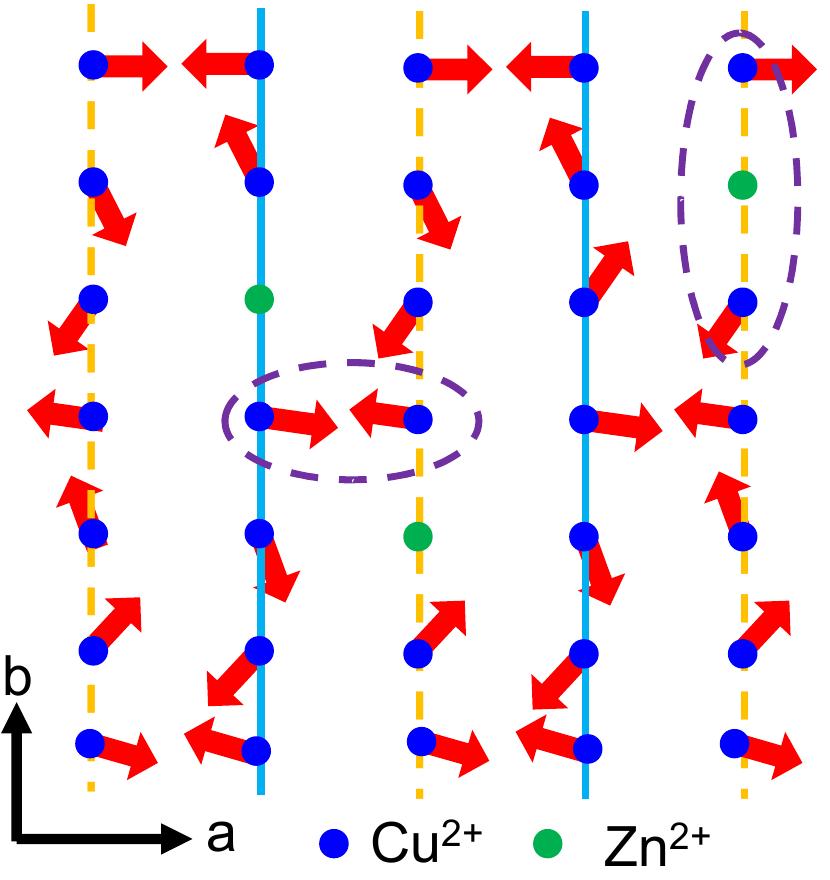}
\end{center}
\caption{\label{fig:fig15}(Color online) A proposed dimer configuration for high Zn substituted ($\geq$ 5.5 $\%$) sample below the phase transition temperature.  This figure is constructed by randomly distributing Zn on a two-dimensional spiral ordered background with a incommensurate 2\textbf{a}$\times$6\textbf{b} magnetic superlattice (modulation $\zeta$ = 0.174)  periodicity along the \textbf{b}-axis.  The alternating dashed orange and solid blue colored lines are not in the same plane but bridged by Cu$^+$ as a quasi-two-dimensional bilayer-A as shown in figure \ref{fig:fig01}.}
\end{figure}

%proposed dimer model description

\subsection{Proposed dimer models}

For a system composed of coexisting isolated spin dimers on a short range ordering helical background, figure \ref{fig:fig15} shows two simplified dimer models we proposed to describe the observed gapped experimental results for Zn substitution level higher than 5.5$\%$.  In the first possible scenario, singlet dimers could form due to non-negligible antiferromagnetic interchain coupling \emph{J$_\bot$} (or \emph{J$_W$}, see figure \ref{fig:fig01}) for interchain coupled spins near Zn pairs.  Effective antiferromagnetic interaction between spins near Zn impurity located on different chains can also be nonzero for coupled frustrated quantum spin chains, where four-spin coupling (cyclic exchange) has been considered.\cite{Laflorencie2005, Laflorencie2003}  Since the ferromagnetic \emph{J$_1$} is disrupted locally by Zn impurity and the antiferromagnetic \emph{J$_2$} becomes dominant locally, the second possible scenario can be described by dimers formed with antiferromagnetic coupled Cu$^{2+}$ ions (\emph{J$_2$}) across the Zn impurities within the same chain (see figure \ref{fig:fig15}).  Such gapped excitation described by a confined spin 1/2 soliton has various gap sizes as spin chain is cut short from the gapless infinite system to various finite lengths.\cite{Furukawa2008}  The distribution of gapped excitations can also be reflected on the non-vanishing spin gap, broad \emph{C$_p$} peaks and $^7$Li NMR 1/T$_1$.  Isolated spin dimers or aggregated clusters coexist on the short-range ordered finite spin chain background when Zn level is above $\sim$ 5.5 $\%$ below \emph{T$_c$}.  The proximity of \emph{T$_c$} defined by \emph{$\chi$(T)} reduction onset and the broad short range ordering signature in \emph{C$_p$(T)}, as described in figure \ref{fig:fig13}, suggest that the coupling strength of short range ordering background spins is comparable or even prerequisite to the isolated spin dimerization near Zn pairs.

For Zn level below $\sim$ 5.5$\%$, dilute isolated spins introduced by Zn should distribute randomly near doped Zn centers.   However, the observed finite size scaling behavior discussed above seems to suggest a scenario that these isolated spins could "transport" to the domain boundaries encircled the long-range helical ordering spins, while Zn impurity should not be mobile at such low temperatures.  Potential spin transport mechanism could be described as a soliton, which can be formed near Zn center and transport through the frustrating helical ordered spin background.  Such behavior supports a picture of finite size effect for Zn $\leq$ 5.5$\%$ and the short-range ordered dimer cluster formation for Zn higher than 5.5$\%$, both of which require spin transport assisted phase separation on the frustrating background.   Mechanism for creation and propagation of a chiral soliton pair on a spiral ordered spin chain background has been described theoretically.\cite{Furukawa2008}   While nonmagnetic Zn ions break the original infinite spin chain of helical spin ordering, soliton pair formed around Zn centers could be transported through this intrachain gapless excitation mechanism until it hit the next Zn impurity, which could be a starting point on interpreting our findings in this study and a detailed calculation is necessary.

Peierls transition has been one of the most common electronic instability mechanisms for the one-dimensional system. There are two types of spin dimer formation, with or without spin-phonon interaction, i.e., whether there is phonon assisted local structure distortion as singlet dimer forms.  Since the phonon assisted spin-Peierls transition requires translational symmetry breaking, i.e., the lattice doubling which loses translational symmetry to open a gap near the Fermi level, which is a collective phenomenon for the one-dimensional spin system.  Although the magnetic field dependence of \emph{T$_h$} shown in figure \ref{fig:fig07} suggests that the gapped phase has a spin Peierls system character, we have not found evidence of lattice doubling from synchrotron x-ray diffraction results so far.  In fact, it is not expected to be observed based on phase separation model, i.e., the phase fraction of isolated dimers is proportional to the Zn level of $\leq$ 10$\%$ only.   On the other hand, we find that \emph{J$_2$} and \emph{J$_\bot$} are both antiferromagnetic and of comparable strength according to LDA calculation and neutron scattering spin wave analysis.\cite{Gippius2004, Masuda2005}  Spins near Zn pair centers must face the frustrating interaction between antiferromagnetic \emph{J$_2$} and \emph{J$_\bot$}.  Interestingly, several LDA and spin wave analysis models suggest the ratio of antiferromagnetic \emph{J$_\bot$}/\emph{J$_2$} to be near 1/2, another condition that satisfies the Majumdar-Ghosh dimerization without phonon assisted lattice distortion.\cite{Majumdar1969}  Whether the dimerization occurs with or without phonon interaction, a local probe on subtle structure distortion is necessary; however, it would be difficult to verify if only a fraction ($\sim$ \emph{z}) of the spins to demonstrate spin dimerization.

Microscopic phase separation due to extrinsic inhomogeneity can often been observed in samples prepared in powder form or thin film on a substrate.  On the other hand, phase separation of electronic origin could be an intrinsic property in a strongly correlated material system and has attracted studies from many aspects, particularly for superconductor and colossal magnetoresistance (CMR) materials.\cite{Dagotto2005}  Since current finite size effect and spin dimer glass behavior are observed in single crystal sample with randomly distributed zero spin ions on a quantum frustrating helimagnetic background,  the phase separation phenomenon in Zn substituted LiCu$_2$O$_2$ system must be of electronic origin.  In particular, even though the driving force remains to be explored theoretically, separated phases distinguished by different spin order/disorder types are unique and have never been reported before to our knowledge.

It is interesting to note that the phase diagram summarized in figure \ref{fig:fig05} shows high similarity to phase boundary and Zn dependence for the general phase diagram as summarized by Bobroff \textit{et al.} on Zn substituted spin gapped systems.\cite{Bobroff2009}   For the gapped systems including isolated ladders, Haldane or spin-Peierls chains, Zn impurities are proposed to induce extended magnetic moments that freeze at low temperatures.  Zn impurities must introduce similar residue spins as shown in figure \ref{fig:fig15} on the spiral ordered background, which could explain the spin freezing phenomenon observed near $\sim$ 5 K from NMR and magnetic susceptibility measurement results discussed above.

%calculated phase diagram by Furukawa2010
The existence of singlet spin dimer has been predicted from DMRG-type algorithm calculation by Furukawa \textit{et al.} for one-dimensional quantum frustrated spin chain with a Hamiltonian similar to equation \ref{eq:one}, where nontrivial easy plane exchange anisotropy is considered also in order to simulate the asymmetric $^7$Li NMR spectra accurately.\cite{Gippius2004, Furukawa2010}  We note that the high Zn substituted samples with \emph{J$_2$/J$_1$} ratios summarized in table~\ref{tab:tableII} could fall in the range of dimerization according to the phase diagram constructed by easy-plane anisotropy vs. \emph{J$_1$/J$_2$} as proposed by Furukawa \textit{et al.}.  In particular, the \emph{J$_2$/J$_1$} ratio drops significantly as system crosses over the $\sim$5.5$\%$ Zn boundary from the original chiral spin ordering to the partially gapped spin dimerization state.  However, we must point out that the phase diagram proposed by Furukawa \textit{et al.} is based on an infinite chain system calculation while current Zn substituted spin chains are finite of unequal lengths, which could explain why the dimerization does not open the spin gap fully but with a size proportional to the Zn substitution level.

%RVB observed?
Finally, the current results also bring in an interesting connection to the possibility of experimental realization of the long sought resonant valence bond (RVB) state.  For a zigzag spin chain system, it has been shown that the RVB state is the exact ground state at the ratio of \emph{J$_2$/J$_1$}= -1/4.\cite{Hamada1988}  Since figure \ref{fig:fig09} shows that  \emph{J$_1$} remains nearly invariant while \emph{J$_2$} decreases with the increasing Zn, one might view our system is actually approaching this RVB critical point at high Zn level, i.e., Zn level higher than $\sim$5.5$\%$ is near the onset of the RVB state.  The RVB state on a finite spin chain due to spinless Zn substitution has gapped low-lying excitations which give rise to a cusp shaped susceptibility as indicated in figure \ref{fig:fig05}b.  Clearly more theoretical exploration is necessary when interchain coupling and easy-plane anisotropy are considered.

\section{Conclusions}

In conclusion, we have used nonmagnetic Zn substitution to explore the quantum quasi-two-dimensional helimagnet LiCu$_2$O$_2$.  Zn substitution moves the frustrated spin system closer to the quantum critical ferromagnetic boundary near $\alpha$ = -1/4 on the \emph{J$_2$} vs. \emph{J$_1$} phase diagram.  Two kinds of magnetic orderings have been found crossing  the \emph{T-z} phase boundary for \emph{z} near $\sim$ 5.5 $\%$ per CuO$_2$ chain.  Low Zn substitution reduces the spiral ordering transition temperatures significantly, following a finite size scaling law, which implies the existence of two-dimensional helimagnetic domains confined by isolated spins, and these isolated spins could be transported as a soliton through the frustrating spin background to the domain boundaries.  Such spin phase separation is uniquely observed for the first time in a two-dimensional helimagnetic background.  High Zn substitution introduces a gapped phase that shows spin-Peierls character and that the phase fraction is proportional to the Zn substitution level.  The proposed interpretations on phase separation, spin dimerization and spin freezing require more microscopic experimental evidences and muon spin resonance investigation of the title compound is under way.

\section*{Acknowledgment}
FCC acknowledges the support from National Science Council of Taiwan under project number NSC-98-2119-M-002-021. The work at McMaster was supported by NSERC, CFI, and CIFAR.\\

\end{document}